\documentclass[11pt]{article}
\usepackage{jcappub,natbib,float,comment,bm}

\usepackage{graphicx,amsmath,amssymb,color,comment,subfigure}

\newcounter{address}
\def\be{\begin{equation}}
\def\ee{\end{equation}}
\def\bea{\begin{eqnarray}}
\def\eea{\end{eqnarray}}
\def\bear{\begin{eqnarray}}
\def\enar{\end{eqnarray}}
\newcommand{\vs}{\nonumber\\}
\def\ba#1\ea{\begin{align}#1\end{align}}
\def\bg#1\eg{\begin{gather}#1\end{gather}}

\newcommand{\g}{\gamma}
\newcommand{\s}{\sigma}
\newcommand{\gobs}{\gamma^{\rm obs}}
\newcommand{\kobs}{\kappa^{\rm obs}}
\newcommand{\refeq}[1]{Eq.~(\ref{eq:#1})}          
\newcommand{\refeqs}[2]{Eqs.~(\ref{eq:#1})--(\ref{eq:#2})}          
\newcommand{\reffig}[1]{Figure~\ref{fig:#1}}          
\newcommand{\refsec}[1]{Section~\ref{sec:#1}}          
\newcommand{\refapp}[1]{App.~\ref{app:#1}}

\renewcommand{\v}[1]{\mathbf{#1}}
\newcommand{\vx}{\v{x}}

\newcommand{\vy}{\v{y}}

\newcommand{\vr}{\v{r}}
\newcommand{\vk}{\v{k}}

\newcommand{\<}{\langle}
\renewcommand{\>}{\rangle}
\renewcommand{\k}{\kappa}
\renewcommand{\d}{\delta}
\newcommand{\D}{\mathcal{D}}
\newcommand{\vth}{\v{\theta}}

\newcommand{\Om}{\Omega_m}

\newcommand{\rhob}{\bar\rho}
\newcommand{\fNL}{f_{\rm NL}}

\def\M{\mathcal{M}}

\def\P{\mathcal{P}}

\def\O{\mathcal{O}}

\def\nhat{\hat{n}}
\def\vnhat{\hat{\v{n}}}

\begin{document}

\title{Imprint of inflation on galaxy shape correlations}

\author[a]{Fabian Schmidt,}
\affiliation[a]{Max-Planck-Institute for Astrophysics, D-85748 Garching, Germany}

\author[b,c]{Nora Elisa Chisari,}
\affiliation[b]{Department of Astrophysical Sciences, Princeton University,
Princeton, NJ~08544, USA}
\affiliation[c]{Department of Physics, University of Oxford, Oxford, OX1 3RH, United Kingdom}

\author[d,\dagger]{Cora Dvorkin}
\affiliation[d]{Institute for Theory and Computation, Harvard University, 60 Garden St., Cambridge, MA 02138}
\affiliation[\dagger]{Hubble Fellow}

\abstract{
We show that intrinsic (not lensing-induced) correlations between galaxy shapes offer a new probe of primordial non-Gaussianity and inflationary physics which is complementary to galaxy number counts.  Specifically, intrinsic alignment correlations are sensitive to an anisotropic squeezed limit bispectrum of the primordial perturbations.  Such a feature arises in \emph{solid inflation}, as well as more broadly in the presence of light higher spin fields during inflation (as pointed out recently by {\it Arkani-Hamed and Maldacena}).  We present a derivation of the all-sky two-point correlations of intrinsic shapes and number counts in the presence of non-Gaussianity with general angular dependence, and show that a quadrupolar (spin-2) anisotropy leads to the analog in galaxy shapes of the well-known scale-dependent bias induced in number counts by isotropic (spin-0) non-Gaussianity.  Moreover, in the presence of non-zero anisotropic non-Gaussianity, the quadrupole of galaxy shapes becomes sensitive to far superhorizon modes.  
These effects come about because long-wavelength modes induce a local anisotropy in the initial power spectrum, with which galaxies will correlate.  We forecast that future imaging surveys could provide constraints on the amplitude of anisotropic non-Gaussianity that are comparable to those from the Cosmic Microwave Background (CMB).  These are complementary as they probe different physical scales.  The constraints, however, depend on the sensitivity of galaxy shapes to the initial conditions which we only roughly estimate from observed tidal alignments.
}

\maketitle
\flushbottom

\section{Introduction}
\label{sec:intro}

The intrinsic alignments of red galaxies are thought to have their origin in the interaction of galaxy shapes with the tidal field of the large-scale structure \cite{Catelan01}. These alignments remain a relatively unexplored cosmological probe, complementary to weak gravitational lensing and the statistics of the galaxy distribution. Upcoming imaging surveys, such as {\it Euclid}\footnote{\url{http://sci.esa.int/euclid/}} and the {\it Large Synoptic Survey Telescope} (LSST\footnote{\url{http://www.lsst.org/lsst/}}), will measure the shapes of billions of galaxies over half of the sky, providing a unique opportunity to exploit alignments as a cosmological probe.

Alignments can contribute to the correlations between shapes and positions of galaxies with a strength larger than weak lensing at low redshifts, and sub-dominant at larger redshifts \cite{Hirata04,Joachimi11}. Much of the effort in the study of alignments has been focused on measuring, modeling and removing intrinsic alignment contamination to weak gravitational lensing observables \cite{Joachimi10,Joachimi10b,Zhang10,Krause15}. The tidal alignment model \cite{Catelan01} has been shown to reproduce the scale-dependence of the observed alignment signal for red galaxies in the linear regime \cite{Blazek11,Singh14} with an amplitude that depends on luminosity \cite{Hirata07}, while blue galaxies have no measured alignments and only upper limits for their intrinsic shape correlations are available \cite{Mandelbaum11,Heymans13}.  In this work, we use ``red'' as a synonym of early-type and ``blue'', of late-type galaxies. In the low-redshift Universe, a very small fraction of early-types have blue colors \cite{Schawinski09}, but this assumption has not been tested at the high redshifts to be probed by upcoming imaging surveys. On the other hand, recent results from numerical hydrodynamical simulations suggest that indeed color is well correlated with galaxy dynamics as tracers of the intrinsic alignment signal \cite{Chisari15b}.  Alignments are typically regarded as a contaminant to `cosmic shear' (the two-point correlation of lensing-induced ellipticities), and they can contribute significantly to the correlation of galaxy shapes with the weak lensing of the Cosmic Microwave Background (CMB) \cite{Hall14,Troxel14}.

The exploration of alignments as a cosmological probe has only started recently. The wealth of shape information expected from upcoming imaging surveys, and the availability of a model that reproduces the observed alignment correlations of red galaxies, has triggered the exploration of the use of intrinsic alignments for cosmology. For example, the correlation between galaxy shapes and CMB $B$-mode polarization induced by primordial gravitational waves could be detected from future surveys, albeit with limitations in confirming its primordial origin \cite{CDS14}. While the CMB $B$-mode polarization is a purely linear effect induced by the gravitational redshift of tensor metric perturbations, the latter also induce an effective tidal field which modifies the local growth of structure and acts to align galaxies \cite{GWshear,tidalpaper}. Another example of the use of intrinsic alignments for cosmology is the possible detection of baryon acoustic oscillations in the cross-correlation of galaxy positions and shapes from future surveys \cite{chisari/dvorkin}. While intrinsic alignments are weaker in amplitude than galaxy clustering, they do have some advantages such as the absence of redshift-space distortions (RSD) \cite{Singh/etal:15}.  

In this work, we show that intrinsic alignments can probe inflationary models where the squeezed-limit bispectrum of the primordial curvature perturbations, usually denoted $\mathcal{R}$ or $\zeta$, is anisotropic.  Phrased in terms of the primordial potential perturbation $\phi$ during matter domination, which is related to the curvature perturbation by $\mathcal{R}=(5/3)\phi$, the squeezed-limit bispectrum can be expressed as 
\be
B_\phi(\vk_1,\vk_2,\vk_3=\vk_L) = \sum_{\ell=0,2,...} A_\ell \P_\ell(\hat\vk_L\cdot\hat\vk_S) 
\left(\frac{k_L}{k_S}\right)^\Delta P_\phi(k_L) P_\phi(k_S)\left[1 + \O\left(\frac{k_L^2}{k_S^2}\right)\right]\,,
\label{eq:BsqI}
\ee
where $k_3 = k_L \ll k_1,\,k_2$ while $\vk_S = \vk_1 -\vk_L/2$, $\P_\ell$ are the Legendre polynomials, and $A_\ell$ are dimensionless amplitudes. Note that the angular dependence has to be even; see \refsec{NG}. In general, there can be several contributions with different power indices $\Delta$. In this paper, we will focus on the ``local'' scaling with $\Delta=0$.  However, our results are easily generalized to other scalings. In \refeq{BsqI}, the coefficient $A_0$ of the isotropic term is related to the usual local non-Gaussianity parameter $f_{\rm NL}^{\rm loc}$ via $A_0 = 4 f_{\rm NL}^{\rm loc}$.  We are interested in constraining the parameter that governs the leading \emph{anisotropic} contribution, namely the quadrupolar dependence of the bispectrum, $A_2$.  As we will show, $A_2$ does not lead to a significant large-scale effect in galaxy clustering, specifically the two-point function.  It does however lead to the analog of the well-known scale-dependent bias for galaxy \emph{counts}, which scales as $f_{\rm NL}^{\rm loc} (k/a H)^{-2}$, in the two-point statistics of galaxy \emph{shapes}, adding a contribution $\propto A_2 (k/a H)^{-2}$.  Combining galaxy clustering with shapes, we can then constrain both isotropic and anisotropic non-Gaussianity simultaneously.

Thus, alignments offer a golden opportunity to explore a new dimension in the parameter space of inflationary models. Anisotropic non-Gaussianity can arise in several early-Universe scenarios. In \emph{solid inflation} \cite{Endlich13}, inflation is driven by the exponential stretching of an unusual solid. A feature of solid inflation is that it produces large anisotropic non-Gaussianity while $f_{\rm NL}^{\rm loc}$ is small ($A_2 \gg A_0$) \cite{Endlich13}.  Both isotropic and anisotropic non-Gaussianity are produced by primordial curvature perturbations generated by large-scale magnetic fields \cite{ShiraishiB1,ShiraishiB2}. Anisotropic non-Gaussianity can also be produced in inflationary models with a generalized bispectrum from excited Bunch-Davies vacuum \cite{Agullo12}.   
Recently, Arkani-Hamed and Maldacena \cite{Arkani-Hamed:2015bza} pointed out (see also \cite{Chen/Wang:1,Chen/Wang:2,Baumann/Green}) that the squeezed-limit bispectrum of primordial perturbations offers a clean probe of the spectrum of particles with masses of order the Hubble scale and smaller during inflation (see \cite{Dimastro15} for a corresponding study of the tensor-scalar-scalar three-point function).  The index $\Delta$ is related to the mass of the particle, with $\Delta\to 0$ for degrees of freedom much lighter than the Hubble scale.  The order of the angular dependence $L$ encodes the spin of the particles, as the leading contribution of a spin-$s$ particle yields a contribution of order $L=s$ in \refeq{BsqI}.  Thus, our results show that the combination of \emph{galaxy clustering and shapes can probe the presence of scalar (spin 0) and tensor (spin 2) degrees of freedom with $m \lesssim H$ during inflation.}

Of course, this type of primordial non-Gaussianity can also be constrained using the CMB, which yields the currently tightest constraints of $A_2 = 13 \pm 93$ ($1\sigma$) \cite{PlanckNG}.   We stress, however, that galaxy clustering and shapes probe the three-point function in \refeq{BsqI} for small-scale modes that are on significantly smaller scales than the CMB.  Given that the amplitude of non-Gaussianity can well be scale-dependent, the constraints we forecast are to be seen as complementary to the CMB.

Our work is organized as follows. Section \ref{sec:NG} presents the non-Gaussian matter density field and our conventions for parametrizing anisotropic non-Gaussianity through the squeezed-limit bispectrum. The impact of tidal fields and  primordial non-Gaussianity on the shapes and clustering of galaxies is described in Section \ref{sec:IA_renorm}. In Section \ref{sec:Cl}, we derive full sky observables of intrinsic alignments. Section \ref{sec:redgal} summarizes the adopted model for the red galaxy population, based on \cite{Joachimi11}, while we outline the Fisher forecast in \refsec{Fisher}. In Section \ref{sec:results}, we present a forecast of the detectability of anisotropic non-Gaussianity in an LSST-like weak lensing survey. Finally, we discuss our results in Section \ref{sec:discuss}.  

Throughout this paper, we assume the following Planck \cite{Ade:2015} fiducial flat $\Lambda$CDM cosmology: $\Omega_{\rm b}h^2=0.022$, $\Omega_{\rm CDM}h^2=0.12$, $h=0.67$, $\Omega_K=0$, $\mathcal{A}_s=2.2\times10^{-9}$, $n_s=0.9645$, $k_p=0.05$ Mpc$^{-1}$ and we define $\Omega_m=\Omega_b+\Omega_{\rm CDM}$. 

\section{Large-scale non-Gaussianity}
\label{sec:NG}

When computing two-point correlations of large-scale structure in the large-scale limit, the leading relevant non-Gaussian statistic is the squeezed-limit potential bispectrum $B_\phi(\vk_1,\vk_2,\vk_L)$, where $k_L \ll k_1,\,k_2$.  While bispectra are in general non-separable functions of the momenta, they can usually be well approximated by a sum of separable terms \cite{SmithZaldarriaga}.  For simplicity, we will only consider one of those terms, which in the squeezed limit can be written in general as
\be
B_\phi(\vk_L,\vk_S,\vk_S) = f(\vk_L, \vk_S) P_\phi(k_L) P_\phi(k_S)\,,
\label{eq:Bsq}
\ee
with ``beyond-squeezed limit'' terms suppressed by $(k_L/k_S)^2$.  These
are negligible when considering large-scale correlations, and we will not consider them here.  
If the bispectrum is scale-invariant, which we will assume in the following,
then $f(\vk_L,\vk_S) = f(\hat\vk_L\cdot\hat\vk_S,\, k_L/k_S)$.   Further,
one can parametrize the dependence on $k_L/k_S$ as a power law $(k_L/k_S)^\Delta$, so that $f(\vk_L,\vk_S) \to f(\hat\vk_L\cdot\hat\vk_S) \,(k_L/k_S)^\Delta$.  
In the following sections, we will restrict to bispectra of the local-type for simplicity ($\Delta=0$).  The generalization beyond the simplifying assumptions made here is straightforward and will be pointed out below where relevant.

Finally we can decompose the angular dependence $f(\mu)$ into multipole moments, where $\mu=\hat\vk_L\cdot\hat\vk_S$: 
\be
f(\mu) = \sum_{\ell=0,\,2,\,4, \cdots} A_\ell \P_\ell(\mu)\,.
\label{eq:fmult}
\ee
It will become clear below why we neglect odd multipole moments.  This
then leads to \refeq{BsqI}.   
Standard local non-Gaussianity corresponds to $L=0$ and $A_0 = 4 \fNL^{\rm loc}$.  
We are interested in the two lowest order contributions:  a monopole $A_0$, and a quadrupole $A_2$, which corresponds to the anisotropic shape generated, for example, by solid inflation \cite{solidinf} or spin-2 particles \cite{Arkani-Hamed:2015bza}.  Thus, $\fNL^{\rm loc}=0$ or ``no local non-Gaussianity'' will correspond to the statement that $A_0=0$ in \refeq{fmult}, while $A_2=0$ corresponds to a vanishing quadrupole.

To interpret the effect of the bispectrum defined in \refeq{Bsq}, consider a single long-wavelength plane wave potential perturbation $\phi(\vk_L)$. According to \refeq{Bsq}, the local small-scale power spectrum at position $\vx$ is given by
\be
P_\phi\left(\vk_S, \vx|\phi(\vk_L)\right) = \left[1 + 
\sum_{\ell=0,\,2,\,4, \cdots} A_\ell \P_\ell(\mu) \left(\frac{k_L}{k_S}\right)^\Delta\, \phi(\vk_L) e^{i\vk_L\vx}\right] P_\phi(k_S)\,,
\label{eq:Pkloc}
\ee
where $P(k_S)$ is the Gaussian power spectrum.  This can be easily verified by multiplying by $\phi(\vk_L')$, integrating over $d^3\vx$, and taking the expectation value.  Since $\phi$ is a real field, $P_\phi(-\vk_S)=P_\phi(\vk_S)$ has to hold, and \refeq{Pkloc} implies that only even multipoles can contribute to the leading squeezed-limit bispectrum \refeq{Bsq}.

\section{Intrinsic alignments and non-Gaussianity}
\label{sec:IA_renorm}

In this section, we derive the linear relation between \emph{intrinsic} galaxy
shapes and the underlying matter distribution in the presence of primordial non-Gaussianity, including the effect of the anisotropic bispectrum.
Intrinsic alignments can be treated in exactly the same way as galaxy biasing, the only difference is that projected shapes are a spin-2 quantity on the sky, as opposed to the scalar galaxy density.
Thus, our discussion parallels that of \cite{dalal/etal,MacDonald08,Slosar08,PBSpaper} who studied the impact of primordial non-Gaussianity on galaxy clustering.  
We describe galaxy shape statistics in two stages.  First, we provide a description of the statistics of the physical, intrinsic three-dimensional shapes of galaxies, denoted as $g_{ij}$.  We will assume that $g_{ij}$ is related to the trace-free part of the second moment tensor $I_{ij}$ of, for example, the intrinsic emissivity of a given galaxy:  
\be
g_{ij} = \left[{\rm Tr}\: I_{kl}\right]^{-1} \left(I_{ij} - \frac13 \d_{ij} {\rm Tr}\: I_{kl}\right)\,,
\ee
so that $g_{ij}$ is dimensionless. Second, we project the shapes onto the sky plane, thus providing the connection to the observed two-dimensional shapes, and add the contribution due to gravitational lensing.  Since the processes that determine galaxy shapes are intrinsically three-dimensional, we argue that a physical treatment of alignments should proceed in this way, rather than starting with an expansion on the projected galaxy shapes.

The precise location and shape of a given galaxy is determined by the nonlinear distribution of matter within some finite spatial region around the galaxy (along  its past trajectory) described by a typical scale $R_*$. For example, if the dominant mechanism for determining position and shape is gravitational collapse, $R_*$ will be of the order of the Lagrangian radius of the parent halo of the galaxy. Since we are interested in the large-scale statistics of galaxies, let us coarse-grain the galaxy distribution on some scale much larger than $R_*$ (but smaller than the scale of correlations we are interested in).  Then, the only properties of the matter distribution that are relevant to galaxy formation are the matter density $\d = \rho/\bar\rho -1$ and tidal field $K_{ij}$,  coarse-grained over a scale $R_*$ (e.g., \cite{baldauf/etal:2011,CFCpaper}).\footnote{More accurately, the relevant properties are the coarse-grained density and tidal field along the entire past trajectory of the fluid elements. However, this is not relevant at the order up to which we work here. See \cite{MSZ,Angulo15} for a discussion in the context of galaxy bias.}  We define the tidal field via
\be
K_{ij} = \frac{1}{4\pi G\, \bar\rho \, a^2} \left[\partial_i \partial_j - \frac13 \d_{ij} \nabla^2 \right] \Phi = \D_{ij} \d,
\label{eq:tijdef}
\ee
where $\Phi$ is the gravitational potential and
\ba
\D_{ij} \equiv\:& \frac{\partial_i\partial_j}{\nabla^2} -\frac13 \d_{ij}
\,.
\label{eq:Dijdef}
\ea
The quantity $K_{ij}$ is trace-free, so that the six degrees of freedom contained in $\{\d,\,K_{ij}\}$ fully describe the second derivative tensor $\partial_i\partial_j\Phi$. Due to the equivalence principle, an observable such as galaxy shapes cannot depend on $\partial_i \Phi$ or $\Phi$ itself. 

Thus, on large scales, $r \gg R_*$, the statistics of galaxy shapes and number, quantified by the fractional overdensity $\d_n$, can be effectively described through local functions of the coarse-grained $\d$ and $K_{ij}$:
\ba
\d_n(\vx, \tau) =\:& F[ \d(\vx, \tau), K_{kl}(\vx, \tau), \tau] \vs
g_{ij}(\vx, \tau) =\:& G_{ij}[ \d(\vx, \tau), K_{kl}(\vx, \tau), \tau]\,,
\label{eq:localbias}
\ea
where $\tau$ denotes conformal time.  
The leading correction to \refeq{localbias} due to the non-locality of galaxy formation is given by terms of order $R_*^2 \nabla^2 \d$ and $R_*^2 \nabla^2 K_{kl}$, which we will neglect throughout this work. In Fourier space, they lead to a scale-dependent bias $\propto k^2 R_*^2$ which becomes relevant on small scales.  

In \refeq{localbias}, we have ignored the small-scale perturbations which
are also present in the real universe.  This is fine in the case of Gaussian initial conditions, since the small-scale perturbations are not correlated with the long-wavelength perturbations [apart from terms that already appear in \refeq{localbias}].  As we will see, this will not be the case for non-Gaussian initial conditions.

A Taylor expansion of \refeq{localbias} then leads to the usual local bias expansion for the galaxy number density perturbation $\d_n$, i.e.
\be
\d_n(\vx,\tau) = b^n_1(\tau)\, \d(\vx,\tau) + \frac12 b^n_2(\tau)\, \d^2(\vx,\tau) + \frac12 b^n_{t}(\tau)\, (K_{ij})^2(\vx,\tau) + \cdots\,.
\label{eq:dglocal}
\ee
The analogous expansion for $g_{ij}$, up to quadratic order, leads to
\ba
g_{ij}(\vx,\tau) =\:& b^I_1(\tau)\, K_{ij}(\vx,\tau) + \frac12 b_2^I(\tau)\, K_{ij}(\vx,\tau) \d(\vx,\tau) 
+ \frac12 b^I_t(\tau) \left[K_{ik} K^k_{\  j} - \frac13 \d_{ij} (K_{lm})^2 \right](\vx,\tau) \vs
& + \cdots\,.
\label{eq:gijlocal}
\ea
Note that we have allowed for all possible terms consistent with the symmetries of $\d_n$ and $g_{ij}$, and in the absence of preferred directions in the galaxy frame. The coefficient $b^n_i$ corresponds to the $i$-th order response of the mean number density of galaxies to a change in the background density, i.e. $b_i^n = (\bar\rho^i/\bar n_g) \partial^i \bar n_g/\partial\bar\rho^i$. Similarly, $b^I_1$ corresponds to the linear response of the mean shape of galaxies to an external tidal field $K_{ij}$.  Due to a lack of preferred directions in the frame of the galaxy, this response must have the simple form given in \refeq{gijlocal}.  That is, $b^I_1$ is simply a number rather than a tensor.  Similarly, $b^I_2$ corresponds to the response of galaxy shapes to an external tidal field and a simultaneous change in the background density.  Such a term is expected to be present intrinsically, however it is also induced by the weighting by the galaxy number density \cite{Blazek/etal:2015}.  Finally, $b^I_t$ quantifies the quadratic tidal alignment.  We emphasise that \refeqs{dglocal}{gijlocal} are \emph{effective} relations describing galaxy statistics on large scales rather than deterministic relations valid at a given scale.  We have also neglected stochastic terms in the bias relations, which have to be included in general.  
The cubic and higher order terms neglected in \refeqs{dglocal}{gijlocal}, while important on small scales, are not relevant for the discussion here as they do not lead to scale-dependent signatures of primordial non-Gaussianity on large scales.  

Let us now consider the cross-correlation of galaxy shapes with the matter density field on large scales. \refeq{gijlocal} yields
\ba
\< \d(\vx) g_{ij}(\vy)\> =\:& b_1^I \< \d(\vx) K_{ij}(\vy) \>
+ \frac12 b_2^I \left\< \d(\vx) \d(\vy) K_{ij}(\vy) \right\>  \vs
& + \frac12 b^I_t(\tau) \left\< \d(\vx) \left[K_{ik} K^k_{\  j} - \frac13 \d_{ij} (K_{lm})^2 \right](\vy) \right\>
+ \cdots\,,
\label{eq:xidg1}
\ea
where the ellipsis stands for higher order correlators which are not relevant on the large scales we are interested in.  The first term in \refeq{xidg1} is the leading term in the case of Gaussian initial conditions.  It is given, in terms of the matter correlation function $\xi(r)$, by
\be
\< \d(\vx) K_{ij}(\vy) \> = \D_{ij} \xi(|\vx-\vy|)\,.
\ee
The second and third terms in \refeq{xidg1} involving three-point functions are exclusively induced by non-linear evolution in the case of Gaussian initial conditions, in which case they are only relevant on small scales. Hence, the large-scale limit of the matter-shape cross-correlation becomes
\be
\< \d(\vx) g_{ij}(\vy)\> \stackrel{\rm Gaussian~IC}{=} b_1^I \D_{ij} \xi(|\vx-\vy|)\,,
\ee
or equivalently in Fourier space,
\be
\< \d(\vk) g_{ij}(\vk')\> = b_1^I \left(\frac{k_i k_j}{k^2} - \frac13 \d_{ij} \right) P_m(k)\, (2\pi)^3 \d_D(\vk+\vk')\,,
\label{eq:PgGauss}
\ee
where $P_m(k)$ is the matter power spectrum.  
This implies that $b_1^I$ is the analog of the usual linear bias for galaxy density correlations.  \refeq{PgGauss} corresponds to the prediction of the so-called \emph{linear alignment model} \cite{Catelan01}.  This is generically the leading contribution to galaxy shape correlations on large scales, unless $b_1^I$ happens to be very small for a given galaxy sample.

We now consider the case of non-Gaussian initial conditions of the local type, quantified by \refeq{BsqI}. The three-point terms  in \refeq{xidg1} are sensitive to the bispectrum in the squeezed limit, since two fields are evaluated at the same point.  In the following, for simplicity of the presentation we will only focus on the first term $\propto b_2^I$;  the term $\propto b_t^I$ behaves in essentially the same way (see \refapp{calc}), so including it here would add nothing new.  As shown in \refapp{calc}, for a squeezed-limit bispectrum of the form \refeq{BsqI} we have 
\ba
\left\< \d(\vx) \d(\vy) K_{ij}(\vy) \right\>
= \frac25 A_2 \D_{ij} \xi_{\d\phi}(|\vx-\vy|) \< \d^2 \>\,,
\label{eq:3pt}
\ea
where $\xi_{\d\phi}(r)$ is the linear matter-primordial potential cross-correlation.  
Equivalently, in Fourier space,
\ba
\left\< \d(\vk) [\d\, K_{ij} ](\vk') \right\>
= \frac25 A_2 \left(\frac{k_i k_j}{k^2} - \frac13 \d_{ij} \right) 
\M^{-1}(k) P_m(k) \< \d^2 \>\, (2\pi)^3 \d_D(\vk+\vk')\,,
\label{eq:3ptF}
\ea
where $P_m(k)$ is the matter power spectrum in synchronous-comoving gauge, and 
\be
\M(k,z) = \frac23 \frac{k^2 T(k) D(z)}{\Om H_0^2}
\label{eq:Mdef}
\ee
is the linear relation between primordial potential and density, $\d(\vk,z) = \M(k,z) \phi(\vk)$, with $T(k)$ denoting the transfer function 
and $D(z)$ denoting the growth factor normalized to $D(z) = 1/(1+z)$ during matter domination.  
Thus, there is a non-Gaussian contribution to the large-scale shape correlation \emph{if the squeezed-limit bispectrum is anisotropic}, $A_2 \neq 0$.  

However, \refeqs{3pt}{3ptF} involve the variance of the matter density field $\< \d^2\>$, which is formally divergent if we let the smoothing scale go to zero, and hence needs to be regularized by introducing a 
counter-term.  Equivalently, it corresponds to an unphysical dependence of our observable on the arbitrary smoothing scale.   
We see that the counter-term we need to add to the bias expansion of galaxy shapes, \refeq{gijlocal}, is $\D_{ij} \phi$.  This term is not due to gravitational evolution, since it involves the potential itself and does not obey the equivalence principle, but is due to the initial conditions.  Note that
in \refeq{localbias} we have implicitly assumed that
the statistics of the initial small-scale modes which influence local galaxy formation
are the same everywhere.  The squeezed-limit bispectrum of \refeq{BsqI}, however,
precisely describes a modulation of the power spectrum of small-scale 
density fluctuations by long-wavelength potential perturbations.  
This exactly parallels the case of galaxy density correlations in 
the case of local-type primordial non-Gaussianity \cite{PBSpaper}.  
The key difference for the generalized squeezed limit of \refeq{BsqI} is 
that now the local power spectrum becomes anisotropic.  Thus,
in the case of primordial non-Gaussianity with $A_2 \neq 0$ we have to explicitly
take into account the dependence of galaxy shapes on the amplitude and
anisotropy of small-scale fluctuations.  

Consider the following transformation of the \emph{initial} matter power spectrum,
\be
P^{\rm ini}_{m,\alpha}(\vk) = \left(1 + 2 \alpha_{lm} \hat k^l \hat k^m\right) P^{\rm ini}_{m,\rm iso}(k)
\label{eq:Pkaniso}
\ee
where $\alpha_{lm}$ is trace-free and $P^{\rm ini}_{m, \rm iso}$ is the fiducial isotropic power spectrum.  We can define the mean response of galaxy shapes to such an anisotropic initial power spectrum,
\be
b_{\rm NG}^I \equiv \frac{\partial \< g_{ij} \>_\alpha}{\partial\alpha_{ij}}\Big|_{\alpha=0}\,,
\ee
where no summation is to be taken (the absence of other preferred directions
requires $b_{\rm NG}^I$ to be a number rather than a tensor).  
As we show in \refapp{newapp}, allowing for this dependence of galaxy shapes on the anisotropy of the initial power spectrum in \refeq{localbias} provides the correct counter-term $3 b_{\rm NG}^I A_2 \D_{ij} \phi$ to cancel the divergence appearing in \refeqs{3pt}{3ptF} (the same holds for the divergence appearing in the other three-point term $\propto b_t^I$).   
This is not surprising, since according to \refeq{Pkloc} (with $\Delta=0$) a long-wavelength potential perturbation yields a modulation of the type \refeq{Pkaniso} with 
\be
2\alpha^{lm} = A_2 \left(\hat k_L^l \hat k_L^m - \frac13 \d^{lm} \right)\phi(\vk_L) e^{i\vk_L \vx}
= A_2 \D^{lm} \phi_L(\vx)\,.  
\ee
The end result is that the large-scale matter-shape cross-correlation becomes 
\ba
\< \d(\vx) g_{ij}(\vy)\> 
=\:& b_1^I \D_{ij} \xi(|\vx-\vy|) + 3 b_{\rm NG}^I A_2 \D_{ij} \xi_{\d\phi}(|\vx-\vy|)\,,
\label{eq:xidgrenorm}
\ea  
and equivalently in Fourier space
\ba
\< \d(\vk') g_{ij}(\vk)\> =\:& 
\left(\frac{k_i k_j}{k^2} - \frac13 \d_{ij}\right) \bigg\{
b_{1}^I   
+ 3b_{\rm NG}^I  A_2\, \M^{-1}(k)
\bigg\} P_m(k)\, (2\pi)^3 \d_D(\vk+\vk') \,.
\label{eq:Pdgrenorm}
\ea

This result is very similar to the well-known scale-dependent bias induced by local primordial non-Gaussianity in the two-point function of the galaxy \emph{density}.  Specifically, the galaxy density-matter cross-correlation is given by 
\cite{dalal/etal}
\be
\< \d(\vk') \d_n(\vk)\> =  \bigg\{
b^n_{1}  
+ \frac12  b^n_{\rm NG} A_0\, \M^{-1}(k) 
\bigg\} P_m(k)\, (2\pi)^3 \d_D(\vk+\vk') \,,
\label{eq:Pdnrenorm}
\ee
where $b^n_{\rm NG} = \bar n_g^{-1} \partial\bar n_g/\partial\ln \sigma_8$ is defined as the response of the galaxy abundance to
a change in the (isotropic) initial power spectrum amplitude quantified by $\sigma_8$.  For dark matter halos and assuming a universal mass function, we can relate this to the linear density bias via $b^n_{\rm NG} = (b^n_1-1) \d_c$.  

Two facts are noteworthy about \refeqs{Pdgrenorm}{Pdnrenorm}.  First, isotropic local type non-Gaussianity ($A_0 \neq 0; A_2 = 0$) does \emph{not generate any large-scale correlations of galaxy shapes}  at linear order in the non-Gaussianity.  The reason is that, due to symmetry, an isotropic rescaling of the small-scale fluctuations cannot affect
galaxy shapes.  Second, there is no impact of anisotropic non-Gaussianity ($A_0 = 0; A_2 \neq 0$) on large-scale galaxy density correlations. The number density of tracers depends on the amplitude of small scale fluctuations. It can be shown (\refapp{calc}) that this is not modified by anisotropic non-Gaussianity.  Thus, at the level of two-point functions, only galaxy shapes are affected by anisotropic non-Gaussianity, while the same is true for the galaxy density and isotropic non-Gaussianity. By comparing shape correlations with density correlations we can thus disentangle isotropic from anisotropic shapes of the local type.  

Finally, it is straightforward to generalize \refeq{Pdnrenorm} to different squeezed-limit power scalings $\Delta$.  Specifically,
\ba
\< \d(\vk') g_{ij}(\vk)\> =\:& 
\label{eq:PdgrenormD}
\left(\frac{k_i k_j}{k^2} - \frac13 \d_{ij}\right)
\bigg\{
b_{1}^I  
+ 3b_{01,\Delta}^I  A_2 \left(\frac{k}{k_p}\right)^\Delta\!\! \M^{-1}(k)
\bigg\}  P_m(k)\,
(2\pi)^3 \d_D(\vk+\vk'),
\ea
where $k_p$ is a pivot scale and we have indicated that now $b_{01,\Delta}^I$
depends on $\Delta$ (and $k_p$).  For $\Delta\approx2$, the non-Gaussian contribution is only weakly scale-dependent and thus becomes more difficult to constrain.  

To summarize, the large-scale correlation between matter and galaxy shapes has a Gaussian piece $\propto b_1^I$, where $b_1^I$ represent the linear response of galaxy shapes to a long-wavelength tidal field;  and a non-Gaussian piece, which is proportional to the quadrupolar part of the squeezed-limit bispectrum.  The coefficient of the non-Gaussian term, $b_{\rm NG}^I$, quantifies the linear response of galaxy shapes to an anisotropic power spectrum of initial fluctuations [\refeq{Pkaniso}].   Our derivations in this section represent the generalization to a spin-2 field of the well-known non-Gaussian halo bias \cite{dalal/etal,slosar/etal,PBSpaper}.  

While it is clear that the scale-dependent bias of galaxy shapes has to be present in general, it is difficult to predict a precise number of $b_{\rm NG}^I$ for dark matter halos, let alone galaxies.  If it is very small, then this effect becomes less interesting observationally.  The value of $b_{\rm NG}^I$ could be measured by running N-body simulations with anisotropic initial power spectrum and measuring the mean shapes of halos along the anisotropy axes.   This is beyond the scope of this paper.  In the context of the excursion set \cite{Bond/etal:91} however, one would expect that  $b_{\rm NG}^I$ is comparable to the linear tidal alignment coefficient $b_1^I$.  In both cases the collapse threshold $\nu=\delta_c/\sigma$ becomes direction-dependent, i.e. different along different axes. In the tidal case, this is due to  the faster infall along the steeper potential gradient (e.g., \cite{Bond/Myers:1996}), while an anisotropic power spectrum of small-scale fluctuations will lead to a similar anisotropic $\nu$.  For our forecasts, we will thus assume that $b_{\rm NG}^I \sim b_1^I$.  This should only be understood as an estimate at the order-of-magnitude level however.  That is, while we expect the strongest signature of anisotropic primordial non-Gaussianity to appear for those galaxies which show the strongest linear tidal alignment (i.e. massive elliptical galaxies), the amplitude of the two coefficients could certainly differ by a factor of several.

\begin{figure}[t]
  \centering
\includegraphics[width=0.6\textwidth]{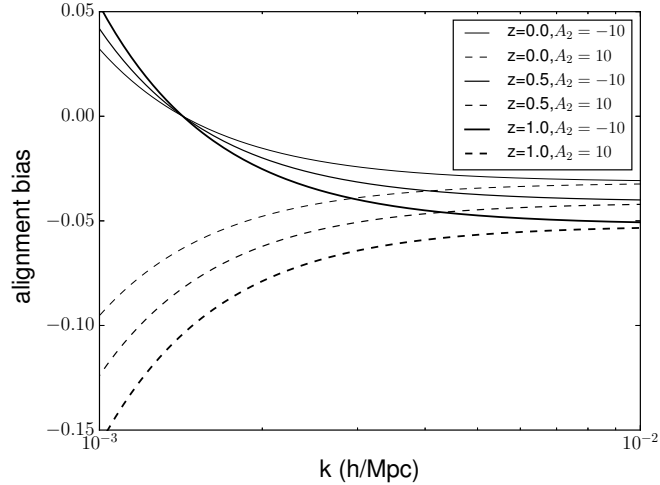}
\caption{The tidal bias induced by anisotropic non-Gaussianity, \refeq{totalbias},
for $b_1^I = - 0.1\Omega_m D(0)/D(z)$, $b_{\rm NG}^I = - 0.1\Omega_m$ and $A_2=-10$ (see \refsec{redgal}). 
The increasing thickness of the curves represents different redshifts from $z=0$ to $z=1.5$. At $k>10^{-2}$ $h$ Mpc$^{-1}$, the effect of non-Gaussianity is negligible and all curves asymptote to $-b_1^I$. Anisotropic non-Gaussianity changes the tidal bias on large scales and the effect is larger at high redshift at a given scale.}
\label{fig:bias}
\end{figure}

Figure \ref{fig:bias} shows the total effective bias of intrinsic alignments, 
\be
 b_1^I  + 3 b_{\rm NG}^I A_2 \M^{-1}(k, z) \,,
\label{eq:totalbias}
\ee
i.e. the term in curly brackets multiplying $P_m(k)$ in \refeq{Pdgrenorm}, as a function of scale.  In the Gaussian case, it is a constant, while for $A_2 \neq 0$ it is strongly scale-dependent.  Here, we assume a fiducial value of $b_1^I = - 0.1\Omega_m D(0)/D(z)$ (consistent with LRG measurements at low redshift \cite{Blazek11}), and $b_{\rm NG}^I= - 0.1\Omega_m$ (see \refsec{redgal}).  The impact of non-Gaussianity increases with redshift. Depending on the sign of $A_2$, the anisotropic tidal bias can have the same sign as the isotropic linear alignment model (if $A_2>0$) or the opposite sign (if $A_2<0$). In the latter case, the effect of the anisotropic tidal bias is to generate {\it tangential alignments} on large scales.  This transition scale can be estimated by simply setting \refeq{totalbias} to zero, and it is $k\simeq 1.5\times10^{-3}$ h/Mpc for $A_2=-10$.  The impact of this bias on angular shape correlations will be shown in the next section.

\section{Observed shape and number count correlations}
\label{sec:Cl}

In this section we derive the implications of the scale-dependent bias
in galaxy alignments for observed angular correlations of galaxy shapes
and number counts.  We will also include the contributions from gravitational
lensing to shapes, which are comparable to or larger than alignments at
cosmological redshifts.  However, we will neglect the impact of primordial
non-Gaussianity on the lensing correlations, since they are highly suppressed
by the projection along the lightcone \cite{shearfNL}.  This is discussed
in detail in \refapp{shearNG}.  

\subsection{Projection}
\label{sec:proj}

In imaging surveys, one does not have access to the 3D shape of galaxies which we have written in \refeq{xidgrenorm}.  Instead, one only measures (in the absence of lensing) the intrinsic shape projected onto the sky:
\be
\g_{I,ij} \equiv \left(\P_i^{\  k} \P_j^{\  l} - \frac12 \P_{ij} \P^{kl}\right)g_{kl} \,,
\label{eq:proj}
\ee
where
\be
\P_{ij} = \d_{ij} - \nhat_i \nhat_j
\ee
is the projection operator and $\vnhat$ is the unit vector along the line of sight. Note that $\P_{ij} \P^{jk} = \P_i^{\  k}$, and any term $\propto \d_{kl}$ in $g_{kl}$ drops out in the projection \refeq{proj}.   
In \refeq{proj} we have introduced the notation $\gamma_{I,ij}$ for the intrinsic shape, as we will also encounter the lensing contribution $\gamma_{G,ij}$ below.  The observed shape $\gamma_{ij}$ is the sum of both contributions.  

Further, in imaging surveys galaxies follow a redshift distribution, $dN/dz$. 
Thus, following \refeq{xidgrenorm}, the intrinsic alignment contribution to observed shapes of galaxies in the direction $\vnhat$ are related to the density field and gravitational potential through
\be
\g_{I,ij}(\vnhat) = \int dz \frac{dN}{dz} \left[ b_1^I \D^\perp_{ij} \d_m(\chi(z)\vnhat; \tau(z)) + 3 b_{\rm NG}^I A_2 \D^\perp_{ij} \phi(\chi(z)\vnhat; \tau(z)) \right]\,,
\label{eq:proj2}
\ee
where $\chi(z)$ is the comoving distance out to redshift $z$, $\tau(z)$ is the conformal time at that redshift, and we have defined the projection onto the sky of the derivative operator $\D_{ij}$ through
\be
\D^\perp_{ij} = \left(\P_i^{\  k} \P_j^{\  l} - \frac12 \P_{ij} \P^{kl}\right) \D_{kl}\,.
\label{eq:Dijperp}
\ee  
Given the discussion of the renormalization approach in \refsec{IA_renorm}, \refeq{proj2} is not to be seen as a truly local relation, but merely as formal  relation in order to derive angular correlations.  

\refeq{Pdgrenorm} shows that the effects of primordial non-Gaussianity on alignment correlations are relevant on large scales.  For this reason, we calculate proper full-sky angular correlations. We begin by reviewing the formalism employed in the following to obtain full-sky correlations efficiently. $\g_{ij}(\vnhat)$ is a traceless 2-tensor on the sky, with two degrees of freedom. Following the literature developed for CMB polarization \cite{GoldbergEtal,ZalSel97,Hu2000}, we decompose $\g_{ij}$ into components that transform with a spin $s=\pm 2$. Consider an orthonormal coordinate system $(\v{e}_1,\v{e}_2,\vnhat)$. If we rotate the coordinate system around $\vnhat$ by an angle $\psi$, so that $\v{e}_i \to \v{e}'_i$, the linear combinations $\v{m}_\pm \equiv (\v{e}_1\mp i\,\v{e}_2)/\sqrt{2}$ transform as 
\be
\v{m}_\pm \to \v{m}'_\pm = e^{\pm i\psi} \v{m}_\pm.
\label{eq:spin1}
\ee
A general function $f(\vnhat)$ is spin-$s$ if it transforms under the same transformation as
\be
f(\vnhat) \to f(\vnhat)' = e^{i s\psi} f(\vnhat).
\ee
An ordinary scalar function on the sphere is clearly spin-$0$, while the unit vectors $\v{m}_\pm$ defined above are spin$\pm1$ fields. This decomposition is particularly useful for deriving multipole coefficients and angular power spectra. For a summary of useful results, see App.~A of \cite{stdruler}.  
We also define
\ba
k_\pm \equiv\:& m_\mp^i k_i \vs
E_\pm \equiv\:& m_\mp^i m_\mp^j E_{ij}
\label{eq:Xpm}
\ea
for any 3-vector $k_i$ and 3-tensor $E_{ij}$.  We then decompose $\gamma_{ij}$ (this equally applies to the individual $I$ and $G$ contributions) as
\ba
\g_{ij} =\:& {}_2\g\, m_+^i m_+^j + {}_{-2}\g\, m_-^i m_-^j \vs
{}_{\pm 2}\g =\:&  m_\mp^i m_\mp^j \g_{ij}\,,
\label{eq:sheardecomp}
\ea
where ${}_{\pm2}\g$ are spin $\pm2$ functions on the sphere (in analogy to Stokes parameters $Q \pm i U$). Consider
the contribution of a single Fourier mode $\vk$ of the density and potential, which we choose along the $z$ axis, to ${}_{\pm 2}\g_I(\vnhat)$:
\ba
{}_{\pm 2}\g_I(\vnhat, k \v{\hat{z}}) =\:& \int dz \frac{dN}{dz} 
\left(\frac{k_\pm}{k}\right)^2\left[ b_1^I  \d_m(\vk; \tau(z)) + 3 b_{\rm NG}^I A_2 \phi(\vk; \tau(z)) \right] e^{i\vk\cdot \vnhat \chi(z)}
\vs
=\:& - \frac12 (1-\mu^2) \int dz \frac{dN}{dz} 
\left[ b_1^I  \d_m(\vk; \tau(z)) + 3 b_{\rm NG}^I A_2 \phi(\vk; \tau(z)) \right] e^{i x \mu}
\,,
\ea
where $\mu$ is the cosine between $\hat{\vk}$ and $\vnhat$, and we have defined $x \equiv k\chi(z)$.  We have also used that $\P_{ij} m_\pm^i m_\pm^j = 0$. 
Since ${}_{\pm 2}\g$ are not scalar quantities on the sky, they do not transform 
trivially under rotations.  Instead, we apply spin-raising $\eth^2$ and lowering $\bar\eth^2$ operators to obtain a scalar on the sky (see App.~A of \cite{stdruler} for details).  Specifically, for a spin-2 quantity such as ${}_{+2}\gamma$, the spin-lowering operator is given by [see Eq.~(A8) of \cite{stdruler} for $m=0$]
\ba
\bar\eth^2\:{}_{+2}\g_I =\:& \frac{\partial^2}{\partial\mu^2}
\left[ (1-\mu^2) {}_{+2}\g_I(\vnhat, \v{\hat{z}})\right] \vs
=\:& \frac12 \int dz \frac{dN}{dz} 
\left[ b_1^I  \d_m(\vk; \tau(z)) + 3 b_{\rm NG}^I A_2 \phi(\vk; \tau(z)) \right] 
(1+\partial_x^2)^2 \left[x^2 e^{i x \mu}\right]\,.
\label{eq:gijk}
\ea
In the last line, we have used
\ba
\frac{\partial^2}{\partial\mu^2}\left[(1-\mu^2)^2 e^{ix\mu}\right]
=& \frac{\partial^2}{\partial\mu^2}\left[(1+\partial_x^2)^2 e^{ix\mu}\right]
= - (1+\partial_x^2)^2 \left[ x^2 e^{ix\mu}\right]\,.
\ea
This derivative operator is real, which reflects the fact that the linear alignments from scalar perturbations considered here do not produce $B$-mode shape correlations even in the presence of anisotropic non-Gaussianity.  $B$-modes will be generated on small scales by the nonlinear alignment terms written in \refeq{gijlocal}, however.  
Applying the spin-raising operator $\eth^2$ to ${}_{-2}\g_I$ yields exactly the same result.  
Crucially, $\bar\eth^2\:{}_{+2}\g_I$ as well as $\eth^2 {}_{-2}\g_I$ transform as scalars on the sky, which
allows us to evaluate their two-point function (angular power spectrum) in a straightforward way.  
The angular power spectrum of $\bar\eth^2\:{}_{+2}\g_I$ is then directly proportional to that of $\g_I$, which is our final goal (see \refsec{corr}).

We emphasize that \refeq{gijk} is the exact, proper relativistic expression of intrinsic shapes on linear scales.  This is because, as shown in \cite{CFCpaper}, the tidal field defined in conformal-Newtonian gauge corresponds to the physical, locally observable tidal field at linear order.  The lensing contribution discussed in \refsec{lens} below then corresponds to the mapping of the intrinsic locally observable shape to our observations from Earth.  Note also that redshift-space distortions do not affect alignments at linear order, since at that order they only modify the physical volume which does not affect shapes (see also \cite{Singh/etal:15}).

Correspondingly, we can also write down the observed galaxy overdensity at linear order:
\ba
\d_n(\vnhat, k \hat{\v{z}}) 
=\:& \int dz \frac{dN}{dz} 
\left[ b^n_1 \d_m(\vk; \tau(z)) + \frac12 b^n_{\rm NG} A_0 \phi(\vk; \tau(z)) \right] e^{i x \mu}\,.
\label{eq:dn}
\ea
While the expression for the intrinsic shapes \refeq{gijk} is exact in the linear regime, 
we have neglected several linear-order contributions in the galaxy overdensity:
\begin{itemize}
\item \emph{Redshift-space distortions:} We will consider photometric galaxy samples throughout, for which RSD are a subdominant effect.  Further, at $l=2$ it is inconsistent to only include the linear RSD term without further relativistic corrections (see below).
\item \emph{Magnification bias:} This contribution, while straightforward to include, is typically subdominant and furthermore affected negligibly by primordial non-Gaussianity as shown in \cite{shearfNL} (see also \refapp{shearNG}).
\item \emph{Relativistic corrections:}  There are corrections to the galaxy clustering kernels (e.g., \cite{challinor/lewis} and App.~B in \cite{GWpaper}) that scale as $(aH/k)$ and $(aH/k)^2$.  The latter terms are of the same form as local-type non-Gaussianity with $A_0 \lesssim 1$ for typical sample parameters.
\end{itemize}
These simplifications are justifiable, since the focus of this paper is shape correlations rather than clustering, 
and since these approximations are not expected to have a significant 
impact on the forecasted constraints for $A_2$.

\subsection{Intrinsic shape correlations}
\label{sec:corr}

We can now apply the results of App.~A1 of \cite{stdruler} to obtain the angular power spectra of galaxy overdensity and shapes, and their cross-spectrum.  
In the following, we will consider three different redshift distributions of galaxies in order to keep the treatment general: $dN_G/dz$ denotes the redshift distribution of all galaxies with measured shapes; $dN_{\rm red}/dz$ denotes the subset of these galaxies that are early-type and known to show linear alignment of the type considered here (a linear alignment has so far not been detected for late-type galaxies).  Finally, $dN_n/dz$ denotes the galaxy sample used in the measurement of the clustering.  For our main results, we will assume that $N_n$, $N_{\rm red}$ and $N_G$ are the same population of red galaxies.  We summarize our model for this population in Section \ref{sec:redgal}.

\refeq{gijk} is a special case, $r=0$, of Eq.~(A19) in \cite{stdruler}, and Eq.~(A24) there (for $|s|=2$ and $N_P=1$, where $N_P$ denotes the number of polarization states) immediately yields the auto-power spectrum of the intrinsic component of the shapes:
\ba
C_I(l) =\:& \frac2\pi \frac{(l-2)!}{(l+2)!} \int k^2 dk\:P_m(k) |F_l^I(k)|^2 \label{eq:ClEE}\\
F_l^I(k) =\:& \int dz \frac{dN_{\rm red}}{dz} 
\left[ b_1^I + 3 b_{\rm NG}^I A_2 \M^{-1}(k, z) \right] \frac{D(z)}{D(0)}
\left[(1+\partial_x^2)^2\: x^2 j_l(x)\right]_{x = k \chi(z)} \vs
=\:& \frac{(l+2)!}{(l-2)!} \int dz \frac{dN_{\rm red}}{dz} 
\left[ b_1^I  + 3 b_{\rm NG}^I A_2 \M^{-1}(k, z) \right] \frac{D(z)}{D(0)}
\left[\frac{j_l(x)}{x^2}\right]_{x = k \chi(z)} \,.
\label{eq:FlE}
\ea
Here, $P_m(k)$ is the matter power spectrum at $z=0$, and $\M(k,z)$ is defined in \refeq{Mdef}.  $j_l$ denotes the spherical Bessel function of order $l$.  
Note that no $B$-modes are generated by linear scalar contributions.  Also,
by symmetry the lowest non-zero multipole of shape correlations is the quadrupole $l=2$.  

Similarly, we can write down the cross-correlation between galaxy overdensity and shape $E$-modes (again, only the intrinsic part). This is the multipole space version of the density-tangential shape correlation function in real space, 
and reads
\ba
C_{nI}(l) =\:& \frac2\pi \sqrt{\frac{(l-2)!}{(l+2)!}} \int k^2 dk\:P_m(k) F_l^I(k) F_l^n(k) \label{eq:ClgE}\\
F_l^n(k) =\:& \int dz \frac{dN_n}{dz} \frac{D(z)}{D(0)}
\left[ b^n_1  + \frac12 b^n_{\rm NG} A_0 \M^{-1}(k, z) \right] 
j_l(x)\Big|_{x=k \chi(z)}\,.
\ea
The clustering auto-power spectrum, on the other hand, is simply given by
\ba
C_{nn}(l) =\:& \frac2\pi \int k^2 dk\:P_m(k) |F_l^n(k)|^2\,.
\label{eq:Clgg}
\ea
\refeqs{ClEE}{Clgg} are valid on the full sky.

\begin{figure}
\centering
\subfigure[$A_2=10$.]{
\includegraphics[width=0.47\textwidth]{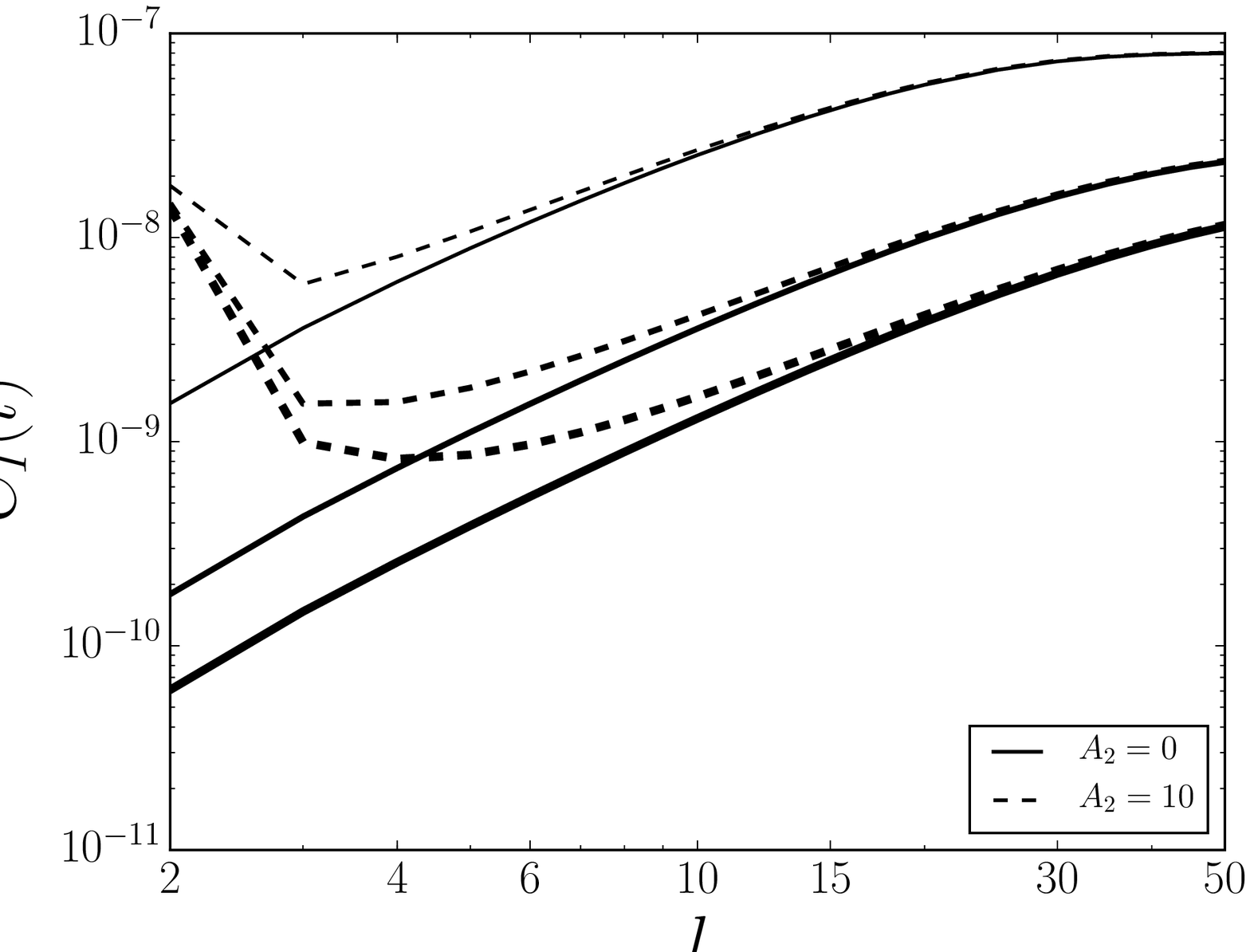}
\label{fig:cleii_a210}
}
\subfigure[$A_2=-10$.]{
\includegraphics[width=0.47\textwidth]{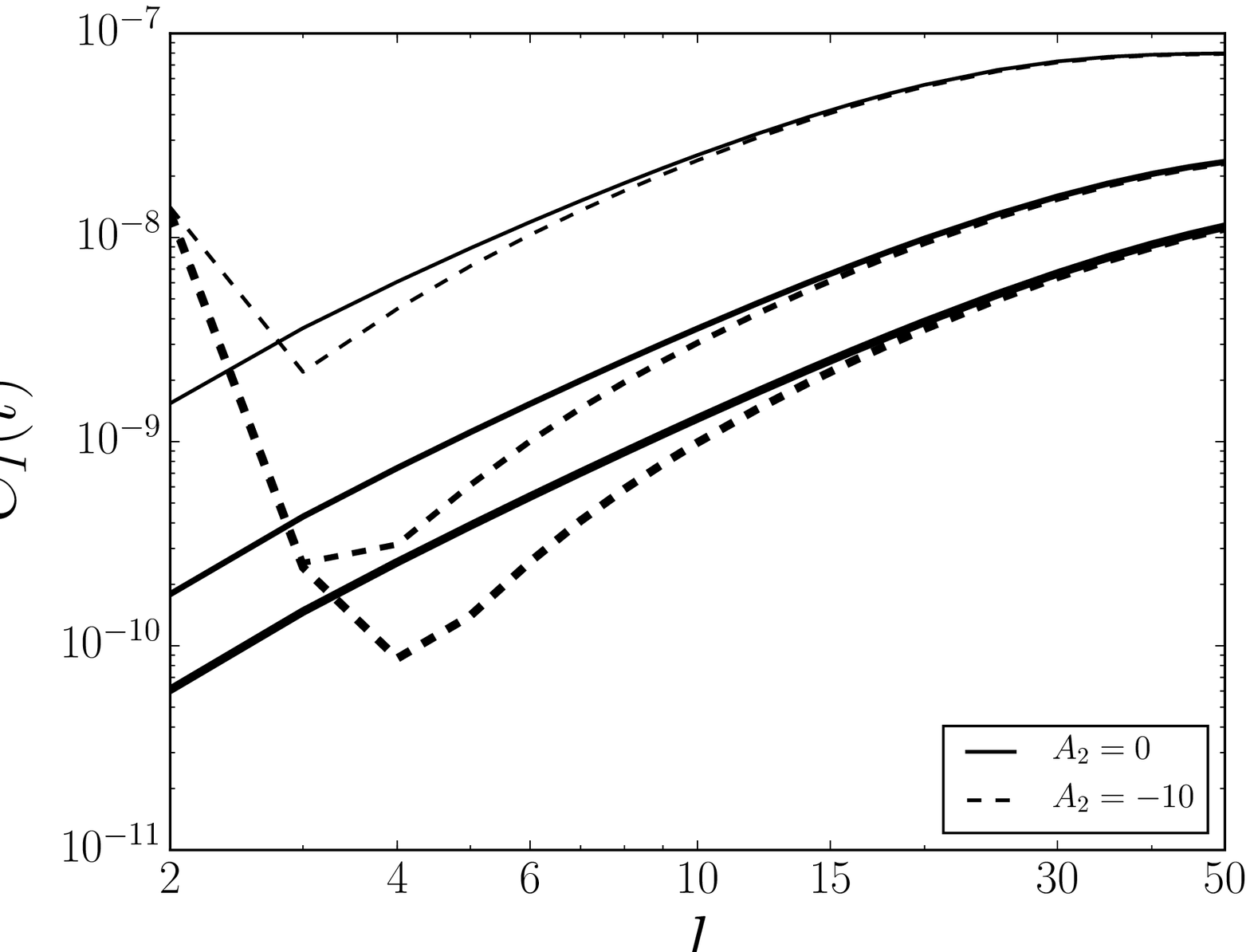}
\label{fig:cleii_a2minus10}
}
\caption{The angular auto-power spectrum of intrinsic shapes (without accounting for weak gravitational lensing) for our fiducial choice of $b_{\rm NG}^I$ and $A_2=+10$ (left) and for $A_2=-10$ (right). The Gaussian case is shown in solid black for three redshifts, $z=\{0.5,1,1.5\}$ from top to bottom. We show the case with anisotropic non-Gaussianity at the same redshifts (dashed). }
\label{fig:cleii}
\end{figure}
\subsection{Signatures of non-Gaussianity} 
\label{sec:psng}

We now illustrate the effect of isotropic and anisotropic non-Gaussianity on the angular power spectra of galaxy counts and the intrinsic contribution to shapes (the lensing contribution will be considered in the next section).  Depending on the sign of $A_0$ and $A_2$, we obtain an enhancement or suppression of the angular correlations at low multipoles (as expected from \reffig{bias}).
In this section, we will assume the same values for the bias parameters as 
for \reffig{bias}, i.e.  $b_1^I = - 0.1\Omega_m D(0)/D(z)$ and $b_{\rm NG}^I= - 0.1\Omega_m$.  

By definition, the intrinsic shape auto-power spectrum (Figure \ref{fig:cleii}) is positive regardless of anisotropic non-Gaussianity. This power spectrum decreases with redshift, as expected from \refeq{ClSS}, from $z=0.5$ (top) to $z=1.5$ (bottom). Anisotropic non-Gaussianity changes the power spectrum for small multipoles ($l\leq 20$) and asymptotes to the Gaussian case on small angular scales. The fractional effect increases towards higher redshifts. 

An interesting and novel effect of primordial non-Gaussianity appears in the quadrupole of shape correlations for $A_2 \neq 0$:  consider the term $\propto (A_2)^2$ in $C_I(l)$.  Focusing just on the $k$-dependence at an effective source redshift $z_s$, this contribution to $C_I(l)$ is proportional to 
\ba
C_I(l) \propto \int k^2 dk\, P_\phi(k) (b^I_{\rm NG} A_2)^2 \left[\frac{j_l(x)}{x^2}\right]_{x= k\chi(z_s)}\,.
\ea
Now consider the quadrupole $l=2$.  We see that the factor in brackets asymptotes to a constant for $k\chi(z_s) \ll 1$, and this term becomes, for $k \ll 1/\chi(z_s)$,
\be
C_I(l) \stackrel{k\chi \ll 1}{\propto} (b^I_{\rm NG} A_2)^2 \int k^2 dk\, P_\phi(k) \,.
\ee
This quantity \emph{diverges logarithmically} in the IR for a near scale-invariant spectrum of primordial fluctuations.  Thus, the quadrupole of galaxy shapes depends on the lowest wavenumber $k_{\rm min}$ at which non-zero anisotropic primordial non-Gaussianity is present, no matter how small.  In principle, a measurement of $C_I(l=2)$, together with the higher $l$ which constrain $A_2$, would thus allow us to measure this low $k$ cut-off, which is presumably related to the beginning of inflation ($k_{\rm min} \sim 1/\tau_{\rm inf}$ where $\tau_{\rm inf}$ is the conformal time at the beginning of inflation).  Quantitatively, the contribution of such super-survey modes is roughly of order $(A_2)^2 \mathcal{A}_s \ln[\tau_{\rm inf}/\chi(z_s)]$, where we have neglected the contribution from the small tilt $n_s-1$.  Given the smallness of the amplitude of fluctuations $\mathcal{A}_s \sim 10^{-10}$, this is probably only observable for an extremely long period of inflation with correspondingly large $\tau_{\rm inf}$ (or, alternatively, if the fluctuations near the begining of inflation happened to be much larger than the extrapolation of the near scale-invariant spectrum).  For definiteness, we choose $k_{\rm min} = 2\times 10^{-8}\, h\,{\rm Mpc}^{-1}$ for our numerical results; for $A_2=10$, $C_I(l=2)$ decreases by roughly a factor of 3 when increasing $k_{\rm min}$ to $10^{-5}\,h\,{\rm Mpc}^{-1}$.  

It might be surprising that modes that are far outside our current horizon affect observables.  Physically, this is because for $A_2\neq 0$, the matter power spectrum in our observable universe is anisotropic and has a spatially constant quadrupole with an expectation value proportional to $(A_2)^2 \int k^2 dk\, P_\phi(k)$.  Galaxy shapes respond to this anisotropy, as quantified by $b^I_{\rm NG}$, and the spatially constant power spectrum quadrupole will lead to a corresponding quadrupole of galaxy shapes.  
Related effects of superhorizon modes have been found in \cite{Dimastrogiovanni14,Schmidt/Hui}.  Note that a similar effect also exists for isotropic non-Gaussianity $A_0 \neq 0$, where the monopole of the galaxy overdensity similarly receives a contribution $\propto (A_0)^2 \int k^2 dk\, P_\phi(k)$.  However, this effect is unobservable, since the amplitude of scalar fluctuations themselves receives the same correction, which is thus absorbed in the normalization $\mathcal{A}_s$.  In this regard, anisotropic PNG and galaxy shapes in principle offer qualitatively new information over the well-known isotropic case.  

\begin{figure}
\centering
\subfigure[Positive $A_2$ combined with $A_0=\{0,10\}$ and compared to the Gaussian case.]{
\includegraphics[width=0.47\textwidth]{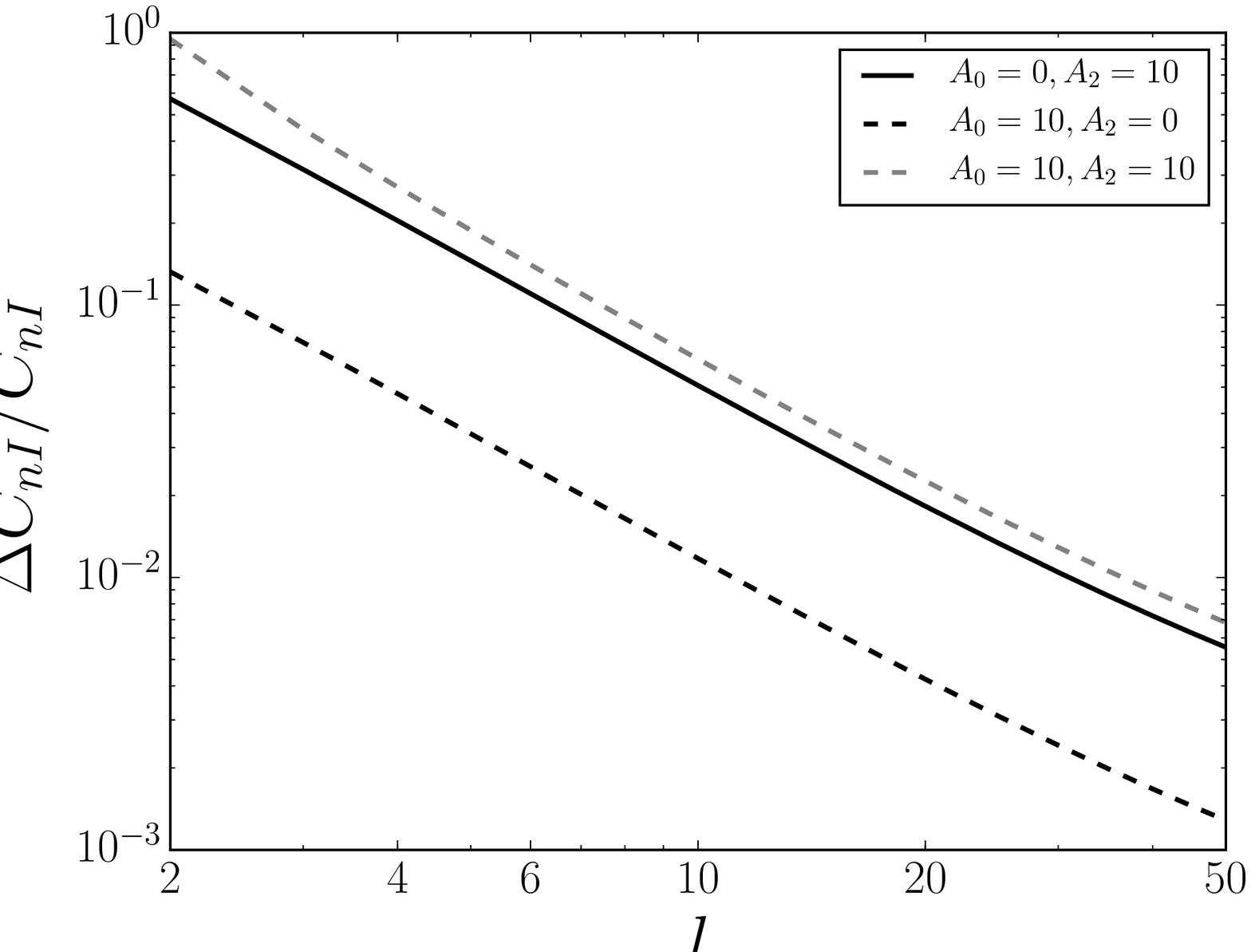}
\label{fig:clegi_a210}
}
\subfigure[Negative $A_2$ combined with $A_0=\{0,10\}$ and compared to the Gaussian case.]{
\includegraphics[width=0.47\textwidth]{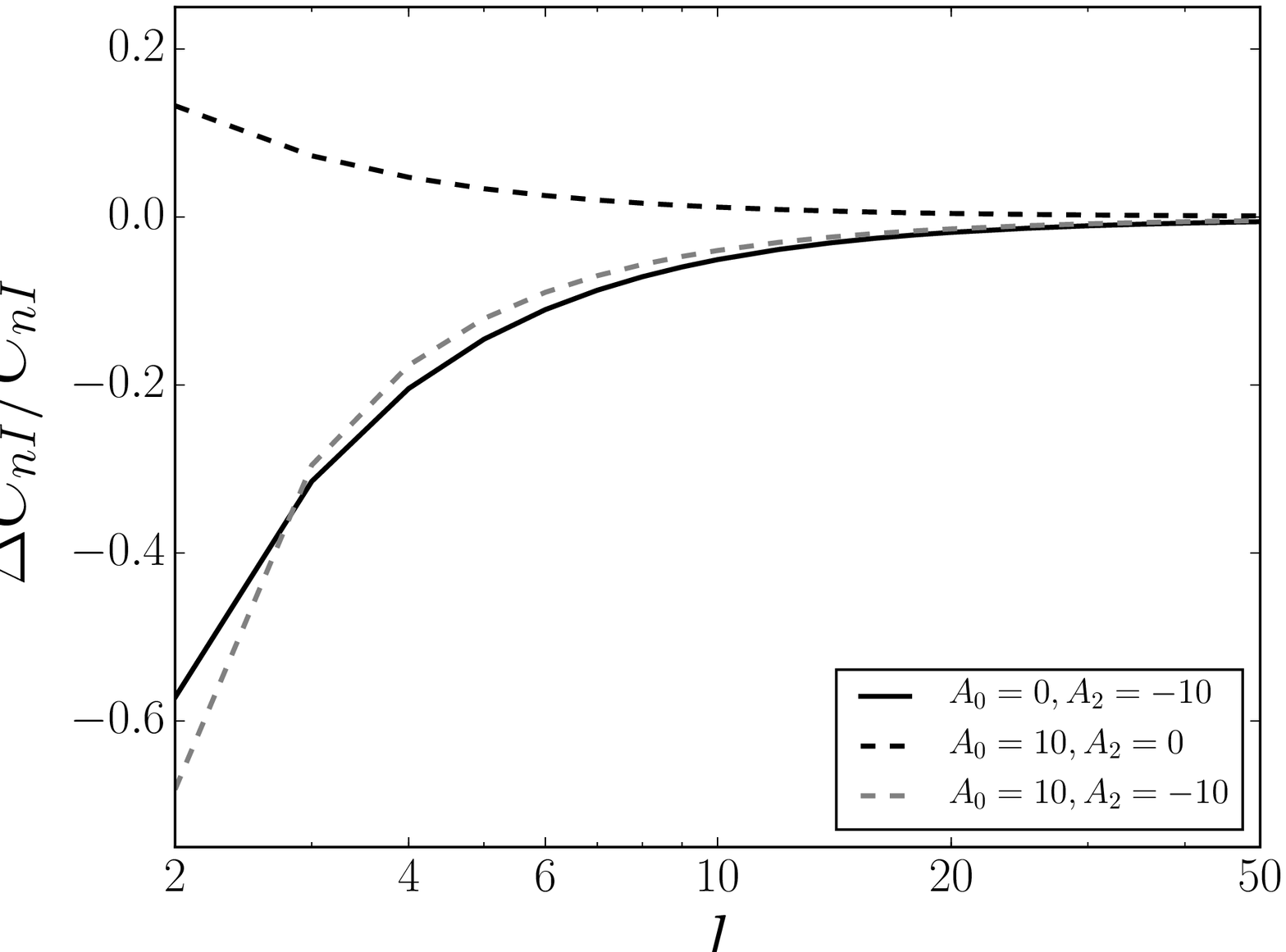}
\label{fig:clegi_a2minus10}
}
\caption{Fractional change in the angular cross-spectrum of galaxy positions and intrinsic shapes at $z=1$ for different levels of non-Gaussianity and our fiducial choice of $b_{\rm NG}^I$. We assume $b_{1}^n=2$ and $b_{\rm NG}^n=(b_{1}^n-1)\delta_c$.  The left panel shows results for $A_2=+10$, while the right panel shows $A_2=-10$.  In addition, we show results for isotropic non-Gaussianity, $A_2=0$, $A_0=+10$ in both panels.}
\label{fig:clgi}
\end{figure}

When the cross-spectrum of galaxy positions and intrinsic shear is considered, both $A_2$ and $A_0$ have an impact on the large-scale correlations. Figure \ref{fig:clgi} shows the effect of the different types of non-Gaussianity on this cross-spectrum and assuming $b_1^n=2$. 
For $A_2 < 0$ (right panel), the correlation between galaxy positions and shapes changes sign on large scales, corresponding to a transition from radial to tangential alignment.  That is, intrinsic shapes in this case have the same sign as the lensing signal, which is also tangential behind overdensities. If $A_2>0$, non-Gaussianity strengthens the effect of the tidal field, enhancing the {\it radial alignment} of red galaxies. In this case, there is no sign change.
We emphasize that both $A_2$ and $A_0$ are theoretically allowed to have either positive or negative signs. 

\subsection{Lensing shear}
\label{sec:lens}

Correlations in galaxy shapes are not only present intrinsically through
tidal alignments, but are also induced by gravitational lensing 
(in particular, the shear; see \cite{Bartelmann01} for a review).  
In general, these two effects are difficult to separate, making it necessary to model them jointly \cite{Joachimi10}. We thus include the effect of gravitational lensing in the shape correlations in our quantitative forecasts of \refsec{results}.  At linear order, the alignment ($I$) and lensing ($G$) contributions are simply additive, so that the full shape auto-power spectrum becomes
\bear
C_{\g\g}(l) &=& \frac{2}{\pi} \frac{(l-2)!}{(l+2)!} \int dk\,k^2\,P_m(k) |F_l^I(k)+F_l^{G}(k)|^2\,,
\label{eq:ClSS}
\enar
where the lensing shear kernel for a distribution of source redshifts $dN_G/dz$ is given by (e.g., App.~F of \cite{stdruler})
\bear
F_l^{G}(k) &=& \frac{1}{2}\frac{(l+2)!}{(l-2)!}  \int_0^{\chi_{\rm max}} d\chi\, k^2\,D_\Phi(k,z(\chi))\,\frac{j_l(x)}{x^2} \chi\int_{\chi}^{\chi_{\rm max}} d\tilde{\chi}  H(\tilde\chi) \frac{dN_G}{dz}\frac{(\tilde{\chi}-\chi)}{\tilde{\chi}} \,,
\label{eq:FlG}
\enar
and $D_\Phi$ is given by
\begin{equation}
D_\Phi(k,z) = \frac{3H_0^2\Omega_{m,0}}{k^2} \frac{(1+z) D(z)}{D(0)}\,.
\end{equation}
The expression in \refeq{ClSS} includes the intrinsic alignment auto-power spectrum, cross-correlations between weak lensing shear and intrinsic shapes, and cosmic shear. The correlation between intrinsic shapes and alignments occur when a galaxy in the background is being lensed by the same matter density field that tends to align the second galaxy \cite{Hirata04,Mandelbaum06}.  
Similarly, the density-observed shape cross-correlation becomes
\ba
C_{n\gamma}(l) =\:& \frac2\pi \sqrt{\frac{(l-2)!}{(l+2)!}} \int k^2 dk\:P_m(k) \left[F_l^I(k) +F_l^G(k)\right] F_l^n(k)\,. \label{eq:ClgE2}
\ea

As this is an important point, we reiterate that the result in \refeq{ClSS}, with \refeq{FlE} and \refeq{FlG}, is exact on linear scales, i.e. it does \emph{not receive further relativistic corrections}.\footnote{This assumes that certain potential gauge artefacts have been removed from $A_2$ \cite{CFCpaper}.}  For the intrinsic contribution, we have argued this in \refsec{proj}.  For the lensing contribution, Ref.~\cite{stdruler} showed that the seemingly ``Newtonian'' expression for the shear in conformal-Newtonian gauge [\refeq{FlG}] is the proper relativistic expression for the observable shear at linear order.  Note that this does not hold for the convergence $\kappa$, where
the observable magnification in fact receives several relativistic corrections.  As discussed above, \refeq{ClgE2} assumes several simplifications in the number count kernel $F_l^n(k)$.  

For the numerical implementation, we use the Limber approximation 
for $l \geq 60$  for all angular power spectra, following \cite{Schmidt08}. 
At these multipoles, the Limber approximation provides an estimation of the angular power spectra with sufficient accuracy for our purposes, and yields better numerical convergence.  We have verified that using the full integrals up to $l=400$ does not change our results.  For the quadrupole ($l=2$), we use a value of $k_{\rm min}=2\times10^{-8}\,h\,{\rm Mpc}^{-1}$ as minimum wavenumber, which matters for the shape auto-correlation as discussed above.  For all other multipoles the value of $k_{\rm min}$ is chosen small enough to ensure a converged result.

\section{Modeling the red galaxy population and alignments}
\label{sec:redgal}

Red galaxies have been observed to be subject to intrinsic alignments with a scale dependence that is well approximated by the linear alignment model (Eq. \ref{eq:PgGauss}) on scales above $\sim $ 10 Mpc$/h$ \cite{Mandelbaum06,Hirata07,Blazek11,Joachimi11,Singh14}. On the other hand, late-type galaxies do not exhibit detectable alignment on large scales \cite{Mandelbaum11,Heymans13}. Thus, in order to estimate the intrinsic alignment power spectrum as a function of redshift, it is necessary to model the red galaxy fraction. 

\begin{figure}
  \centering
\includegraphics[width=0.6\textwidth]{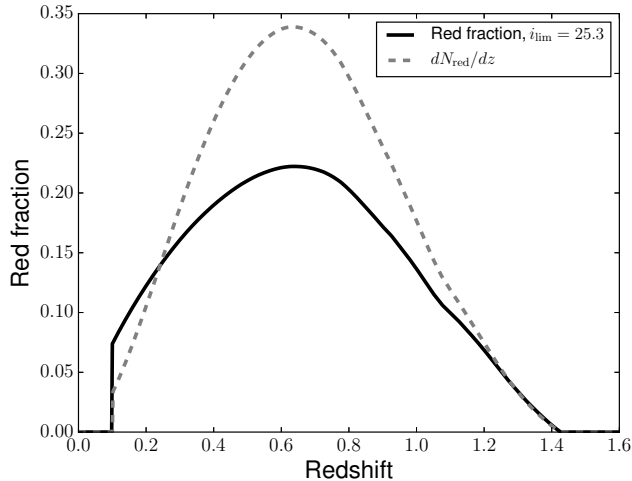}
\caption{The expected fraction of red galaxies observed by LSST as a function of redshift for a limiting magnitude of $i_{\rm lim}=25.3$ (AB). The gray dashed curve shows $dN_{\rm red}/dz$ with an arbitrary normalization.}
\label{fig:lsst_redfrac}
\end{figure}

The fraction of red sources, 
\be
f_{\rm red}(z) = \frac{dN_{\rm red}/dz}{dN_G/dz}\,,
\ee
depends on the intrinsic properties of red galaxies (their spectral energy distribution and luminosity function) and on the properties of the survey (filter, magnitude limit, source redshift distrubution). We use the model constructed by Ref. \cite{Joachimi11} and applied by Ref. \cite{Chisari15a} to obtain the fraction of red galaxies and their mean $i$-band luminosity as a function of redshift for an LSST-like survey. The source redshift distribution expected for LSST, which extends from $z=0.1$ to $z=3$, is given in \cite{Chang13}; there are $26$ galaxies per arcmin$^2$ and we assume an $i$-band limiting magnitude of $25.3$ (AB), consistent with the so called `gold sample' of galaxies with signal-to-noise ratio $>20$ and measured shears in LSST \cite{LSST}. We assume a red galaxy luminosity function as observed by Ref. \cite{Faber07} from a combination of galaxy surveys between $0.2<z<1.2$. In our model, the paucity of red galaxies at $z>1.5$ implies that intrinsic alignments have no contribution to the total shear above those redshifts. The redshift distribution of red galaxies in LSST using this model is shown in Figure \ref{fig:lsst_redfrac}. Throughout, we assume a survey with an area coverage of $18,000$ square degrees, since the effect of masking is accounted for in the effective number density of galaxies. When projected over the range $z=[0,1.5]$, LSST should observe $3$ red galaxies per arcmin$^2$. Notice that this is optimistic because we are not modeling any additional selection cuts on galaxy shapes that could potentially affect the red population. In our model, blue galaxies are assumed to have no alignments.

We can relate the linear shape bias $b_1^I$ to the convention commonly used in papers reporting on measurements of intrinsic alignment as follows:
  \begin{equation}
    b_1^I = - A_I(L_r) \Omega_m C_1 \rho_{\rm crit} \frac{D(0)}{D(z)},    
  \end{equation}
where $\Om$ is the matter density parameter today and $C_1\rho_{\rm crit}=0.0134$ is fixed by convention to the SuperCOSMOS intrinsic alignment amplitude \cite{Brown02}. Later on, we will treat $C_1\rho_{\rm crit}$ as a free parameter when making predictions of non-Gaussianity constraints from future surveys.  
$A_I$ is a luminosity-dependent amplitude measured by Ref.~\cite{Joachimi11} from a sample of Luminous Red Galaxies (LRGs) in the {\it Sloan Digital Sky Survey} \cite{Eisenstein01},
\begin{equation}
A_I(L_r) = (5.76^{+0.60}_{-0.62}) \left( \frac{ L_r }{L_0}\right)^{1.13^{+0.25}_{-0.2}},
\label{eq:AIJ11}
\end{equation}
where $L_r$ is the average $r$-band luminosity at a given redshift and $L_0$ is a pivot luminosity corresponding to an absolute magnitude of $M_0(z)=-22-1.2z-5\log_{10}h$.  We emphasize that ongoing imaging surveys, such as the {\it Kilo-Degree Survey}\footnote{\url{http://kids.strw.leidenuniv.nl}}, the {\it Dark Energy Survey}\footnote{\url{http://www.darkenergysurvey.org}}, the {\it PAU} survey\footnote{``Physics of the Accelerating Universe'', \url{http://www.pausurvey.org/home-PAU.html}} and {\it Hyper-Suprime Cam}\footnote{\url{http://www.naoj.org/Projects/HSC/}} will be able to test the assumptions on the dependence of $b_1^I$ with redshift and luminosity we have made here.

Unlike in the case of the scale-dependent bias in galaxy number counts, $b^n_{\rm NG} \approx (b^n_1 -1) \delta_c$, for which a fairly accurate estimate is known, no such estimate exists for the response of shapes to the power spectrum anisotropy quantified by $b_{\rm NG}^I$.  We will thus assume that $b_{\rm NG}^I$ is comparable to $b_1^I$ at any given redshift, specifically
\begin{equation}
  b_{\rm NG}^I = \tilde{b}_{\rm NG}^I \times b_1^I(z) \frac{D(z)}{D(0)}
\end{equation}
where we assume throughout $\tilde{b}_{\rm NG}^I=1$ but keep it explicit in order
to illustrate the residual modeling uncertainty.

Note that the effect of the luminosity dependence of the intrinsic alignment bias was not taken into account in \refsec{psng} above, but will be included in the quantitative forecast.  Its most important consequence is the existence of a Malmquist-like bias for the population of intrinsically aligned galaxies in a magnitude-limited survey. As a result, the red galaxy population at higher redshifts has a larger $b_1^I$, which results in a larger contribution from anisotropic non-Gaussianity compared to the case presented in \refsec{psng}, where $b_1^I$ was set to a constant.

\section{Fisher matrix analysis}
\label{sec:Fisher}

In order to estimate the ability of future surveys to set joint constraints on 
isotropic and anisotropic non-Gaussianity, 
we perform a Fisher matrix analysis over the parameter set $\{b_1^n,A_0,A_2,C_1\rho_{\rm crit},\sigma_8\}$.  The remainder of the cosmological parameters are fixed to their values given at the end of \refsec{intro}.  
Further, we include a prior in $\sigma_8=0.831\pm0.013$ from Planck \cite[][Table 3, column 4]{Ade:2015}.

The predictions for the angular auto- and cross-power spectra were given in \refsec{Cl}.  All auto-power spectra also contain noise terms, which we assume to be white,
\be
N^{\gamma} = \frac{\sigma_\gamma^2}{\bar{n}_G}
\quad\mbox{and}\quad
N^n= \frac1{\bar{n}_n}\,,
\ee
where $\sigma_\gamma^2$ is the dispersion of the intrinsic shape including measurement noise per component, and $\bar{n}_G$ is the projected surface density of galaxies with shapes per steradian.  Similarly, $\bar{n}_n$ is the surface density of galaxies used to compute the clustering signal. 

Since all perturbations are Gaussian at the order we work in, the covariance between two power spectra is given by 

\begin{equation}
  {\rm Cov}[C_{\alpha\beta}(l),C_{\gamma\delta}(l')]  = \delta^{ll'}_K \frac{1}{(2l+1)f_{\rm sky}}\left[C_{\alpha\gamma}(l)C_{\beta\delta}(l)+C_{\alpha\delta}(l)C_{\beta\gamma}(l)\right]
\end{equation}
where the indices $\{\alpha,\beta,\gamma,\delta\}$ run over $\{n, G\}$, i.e. number densities and shapes.

The Fisher matrix is given by \citep{Tegmark97,Duncan14}

\begin{equation}
F_{\mu\nu} = \sum_l \frac{\partial {\bf D}(l)}{\partial p_{\mu}} {\rm Cov}^{-1}(l)  \frac{\partial {\bf D}(l)}{\partial p_{\nu}}
\end{equation}
where ${\rm Cov}$ is the covariance matrix of the data vector ${\bf D}=\{C_{nn},C_{n\gamma},C_{\gamma\gamma}\}$ and $\partial_\mu$ indicates the partial derivative with respect to the parameters of interest.  We perform these derivatives analytically on the expressions for $C_{\alpha\beta}(l)$, and evaluate the results numerically.  In practice, $A_2$ cannot be constrained independently of $\tilde{b}_{\rm NG}^I$. While we assume $\tilde{b}_{\rm NG}^I=1$, this means that the power spectra are effectively constraining the product $\tilde{b}_{\rm NG}^IA_2$. The $1\sigma$ uncertainty in each parameter, after marginalizing over all other parameters, are given by $\sigma_{\mu,{\rm full-margin}}=\sqrt{(F^{-1})_{\mu\mu}}$. We also consider forecasted errors after marginalizing only over $\{b_1^n,C_1\rho_{\rm crit},\sigma_8\}$, which allows us to determine whether degeneracies between $A_0$ and $A_2$ exist. If $A_0$ and $A_2$ correspond to coordinates $1$ and $2$ of the Fisher matrix, we obtain the partially-marginalized bounds from defining the following sub-matrix,
\begin{equation}
(G^{-1})_{i,j}=[F^{-1}]_{i=[1,2],j=[1,2]}
\end{equation}
and then the partially-marginalized uncertainties are given by
\begin{equation}
  \sigma_{i,{\rm part-margin}}=G_{ii}^{-1/2}.
\end{equation}

\section{Constraining non-Gaussianity with clustering and shapes}
\label{sec:results}

\begin{figure}
  \centering
  \subfigure[$C_{n\gamma}(l)$]{
    \includegraphics[width=0.45\textwidth]{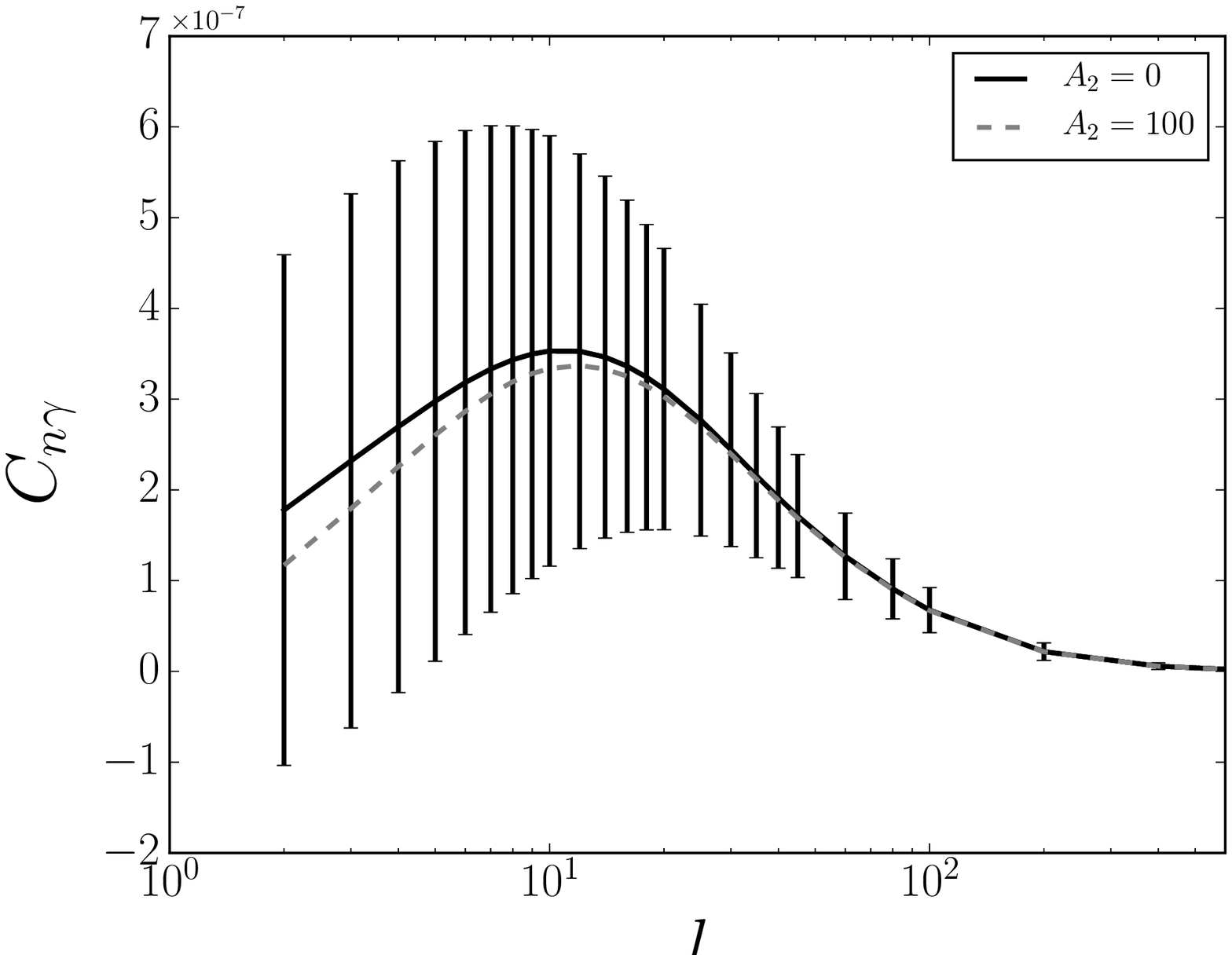}
  }
  \subfigure[$C_{\gamma\gamma}(l)$]{
    \includegraphics[width=0.45\textwidth]{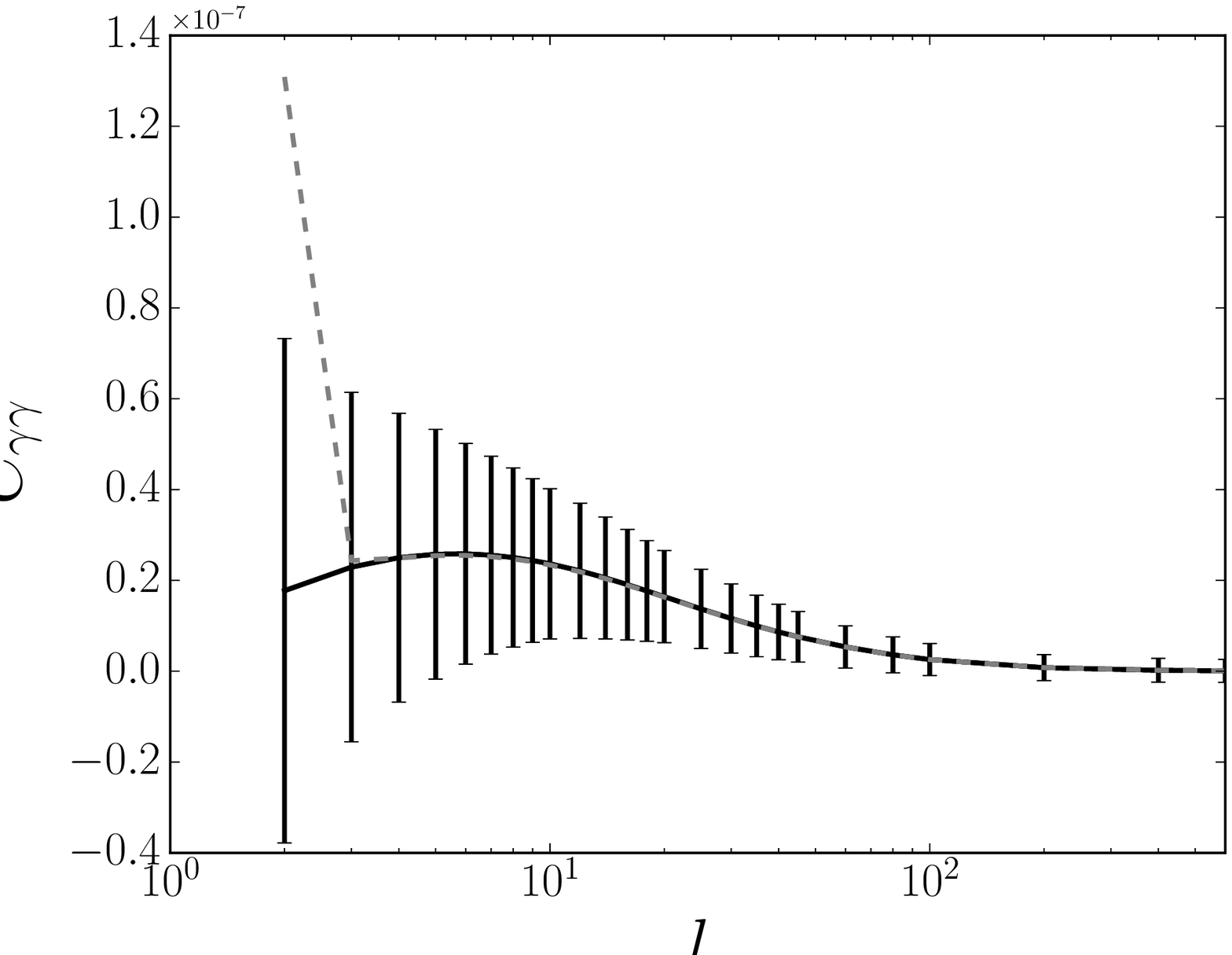}
    }
\caption{Position-shape correlations and shape-shape correlations of red galaxies expected to be observed in LSST between $0.1<z<1.5$. The case of Gaussian initial conditons is shown in solid black, and the case with $A_2=100$ is shown as dashed gray lines. The error bars show the expected uncertainties for a Gaussian fiducial cosmology.}
\label{fig:clswerr}
\end{figure}

In this section, we forecast the ability of galaxy imaging surveys to detect anisotropic non-Gaussianity using galaxy shapes, in particular those of red galaxies.  We consider an LSST-like survey with $26$ galaxies per arcmin$^2$ \cite{Chang13} and we model the red galaxy population as described in Section \ref{sec:redgal}. We assume that the dispersion in the intrinsic ellipticities per component is $\sigma_\g = 0.3$ \cite{LSSTscience}. We do not account for dispersion due to photometric redshift scatter in the red galaxy population in this section, since we are integrating over the full redshift range of the red galaxy population. Figure \ref{fig:clswerr} shows the position-shape and shape-shape angular power spectra and their expected uncertainties as a function of multipole number for the cases with $\tilde{b}^I_{\rm NG}A_2=0$ and $\tilde{b}^I_{\rm NG}A_2=100$.  Clearly, the shape correlations are dominated by the lensing contribution, and the fractional effect of anisotropic non-Gaussianity is significantly smaller than in the intrinsic-only correlations shown in Figures \ref{fig:cleii} and \ref{fig:clgi}.  

We combine the clustering, galaxy-lensing and cosmic shear information for the red galaxy population to obtain constraints on $\{C_1\rho_{\rm crit}, b_1^n, A_0, (\tilde{b}_{\rm NG}^IA_2),\sigma_8\}$. Notice that we assume that the luminosity and redshift dependence of the intrinsic alignment model is fixed by current observational data, but we allow the overall amplitude of alignments to be constrained through changing the normalization given by $C_1\rho_{\rm crit}$ (no priors are considered). Analogously, we assume a fiducial constant, redshift-independent bias for red galaxies, $b_{1}^n=2$, that is allowed to vary in the Fisher analysis. We obtain the following fully marginalized $1\sigma$ bounds on the uncertainties in the parameter set from the multipole range $2\leq l\leq 600$: 
\begin{eqnarray}
\Delta(C_1\rho_{\rm crit}) &=& 8.6 \times 10^{-4},\nonumber\\ 
\Delta b_1^n &=& 2.6 \times 10^{-2}, \nonumber\\ 
\Delta A_0 &=& 44,\,{\rm and}\nonumber\\ 
\Delta (\tilde{b}_{\rm NG}^IA_2) &=&  150.
\end{eqnarray}
Our forecast for the uncertainty in $A_2$ (assuming $\tilde{b}_{\rm NG}^I=1$) is somewhat worse than the current CMB constraints, $\Delta A_2^{\rm CMB} \sim 90$ \citep[][at $68\%$ confidence level]{Shiraishi13,PlanckNG}.  We emphasize however that galaxy alignments probe different scales than the CMB, and thus these constraints should be seen as complementary to the CMB in view of a possible scale-dependence of the non-Gaussianity.  This applies in particular to the small-scale modes, where the modes relevant for galaxy formation are $k_S \gtrsim 1\,h\,{\rm Mpc}^{-1}$, while the CMB probes modes with $k_S \lesssim 0.1\,h\,{\rm Mpc}^{-1}$.    

Note further that the constraints depend on the fiducial value of $A_2$ chosen, where our choice of fiducial $A_2=0$ is conservative.  This is because the terms $\propto (A_2)^2$ do not contribute to the constraints in this case.  In particular, this applies to the quadrupole of the shape auto-power spectrum, which diverges as $k_{\rm min}\to 0$ as discussed in Section~\ref{sec:psng}, and is the most prominent non-Gaussian feature in \reffig{clswerr}.  Thus, the true log-likelihood is steeper than the quadratic approximation around $A_2=0$ suggests, so that even in case of a non-detection the actual constraints obtained for a survey with these specifications could be tighter.  Further, in case of a detection of non-zero $A_2$ with shape correlations, the amplitude of $C_{\gamma\gamma}(l=2)$ could be used to constrain $k_{\rm min}$ via its logarithmic dependence on it.

\begin{figure}
  \centering
\includegraphics[width=0.6\textwidth]{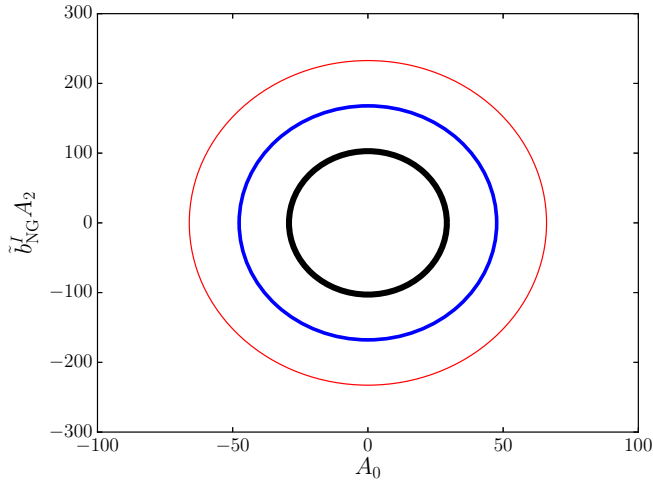}
\caption{$\{1,2,3\}\sigma$ uncertainties in $A_0$ and $\tilde{b}_{\rm NG}^IA_2$ (thick to thin black, blue and red solid lines, respectively) 
marginalizing over $C_1\rho_{\rm crit}$, $\sigma_8$ and $b_1^n$. 
These constraints are obtained from considering the combination of clustering, 
position-shape correlations and shape-shape correlations and using the redshift 
distribution of red galaxies expected to be observed in LSST between $0<z<1.5$.}
\label{fig:ellipses}
\end{figure}

For the well-known isotropic non-Gaussianity on the other hand, our results indicate an uncertainty of $\Delta\fNL^{\rm loc} = 11$.  This latter constraint is significantly worse than what is expected for Stage IV \emph{spectroscopic} surveys, since we consider only a single redshift bin which loses information on all line of sight modes. We will return to this issue at the end of this section.

Figure \ref{fig:ellipses} shows likelihood contours in the  $A_0-\tilde{b}_{\rm NG}^IA_2$ plane, obtained by marginalizing over $C_1\rho_{\rm crit}$, $\sigma_8$ and $b_1^n$.  They correspond to $\{1,2,3\}\sigma$ confidence intervals.  We find that including all auto- and cross-correlations of number counts and shapes almost entirely removes any degeneracy between the two non-Gaussianity parameters.

\begin{figure}
  \centering
  \includegraphics[width=0.6\textwidth]{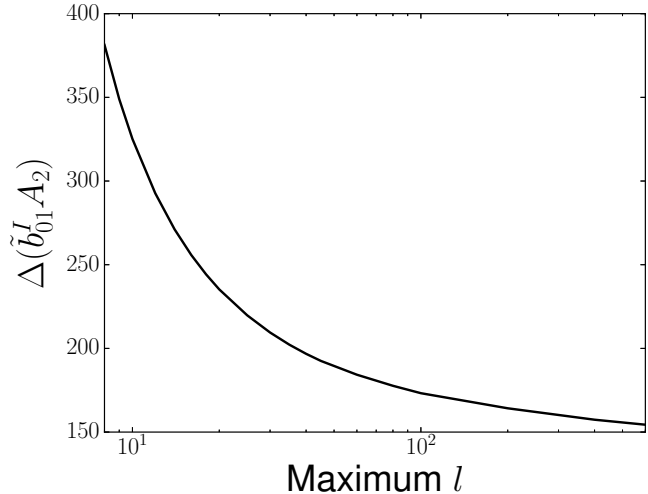}
  \caption{Bound on $\Delta(\tilde{b}_{\rm NG}^IA_2)$ after marginalizing over all other parameters and as a function of the maximum multipole considered (holding the minimum $l$ fixed at $2$).}
  \label{fig:onesigA2}
\end{figure}

Given our linear treatment of the matter power spectrum, we consider the effect of restricting the forecasts to lower multipoles (note however that nonlinear corrections to the lensing contribution are straightforward to incorporate). Figure \ref{fig:onesigA2} shows the forecasted uncertainty in the anisotropic non-Gaussianity parameter as the maximum $l$ is increased. This figure shows that although the effect of non-Gaussianity is most significant on large scales, the small scales still contain significant information in particular on $b_1^n$ and $C_1\rho_{\rm crit}$ which are helpful in improving the constraints on the non-Gaussianity parameters.  While extending the modeling of the power spectra to nonlinear scales is beyond the scope of this work, there is thus potential for better constraints on the non-Gaussianity parameters from $l>600$. However, the effects of baryonic physics on the nonlinear lensing and alignment power spectrum on those scales are still a matter of debate (see, for example, \cite{vanDaalen11}).
On the other hand, removing scales below $l<10$ while keeping $l_{\rm max}=600$ fixed increases the uncertainty in the marginalized $\Delta(\tilde{b}_{\rm NG}^IA_2)$ by a factor of $1.6$. Not surprisingly, the very low multipoles are important for constraining primordial non-Gaussianity, requiring careful control of the systematics in imaging surveys.  

In our constraints we have included the full redshift distribution of galaxies, thus suppressing any information contained in modes along the line of sight.  This is the reason why our forecasted constraint on $f^{\rm loc}_{\rm NL}$ is roughly an order of magnitude worse than what is obtained from spectroscopic surveys of similar volume.  Given that, unlike the lensing field, the tidal alignment is an intrinsically three-dimensional effect, we thus expect a potential significant improvement over the constraints quoted here by using a full tomographic analysis rather than a single redshift bin, which would restore information on the line-of-sight modes.  This has been found to hold for the scale-dependent bias in the galaxy density induced by $f^{\rm loc}_{\rm NL}$ \cite{Cunha/etal,Leistedt/etal}, and for IA in the case of Gaussian initial conditions \cite{Krause15,DES/shear}.  Further, tomography should help break any degeneracy between
the alignment strength and cosmological parameters such as $\sigma_8$, which we have not considered here.  In fact, a much smaller spectroscopic sample of red galaxies with well-measured shapes could deliver interesting constraints as well.  
We defer the forecast for the full tomographic and spectroscopic cases to future work.

\section{Discussion and conclusions}
\label{sec:discuss}

We have shown that the intrinsic alignment of red (elliptical) galaxies will become an interesting probe of inflationary physics in the future.  Through their sensitivity to a quadrupolar anisotropic mode coupling, galaxy shapes expand the parameter space of inflationary models that can be probed using large-scale structure observables.  The anisotropic squeezed-limit bispectrum translates into a local anisotropy in the matter power spectrum, which imprints itself in the shapes of halos and galaxies forming at that location. This non-Gaussian bias produces a tendency towards radial or tangential alignments, depending on the sign of the non-Gaussianity, which increases rapidly towards large scales.  This is the analog in galaxy shapes of the familiar scale-dependent bias in galaxy clustering induced by isotropic primordial non-Gaussianity. 

As a specific example, models such as solid inflation, the realization of inflation through the symmetries of a solid \cite{Endlich13}, could be probed through this method.  
More generally, as was recently pointed out by Ref. \cite{Arkani-Hamed:2015bza}, the angular dependence of the bispectrum encodes information about the spin of light particles during inflation. Therefore, constraints on the quadrupolar part of the bispectrum, analyzed in this work, provide information about tensor degrees of freedom during inflation.  Moreover, we have shown that should anisotropic non-Gaussianity be detected, the quadrupole of galaxy shapes can be used to constrain the \emph{maximum} scale at which this mode coupling persists, which is presumably related to the beginning of or some other special event during inflation.  This is a unique possibility of probing physics outside of our horizon, which is not possible with isotropic non-Gaussianity.

We have modeled the red fraction of galaxies in the LSST survey following the approach suggested by \cite{Joachimi11}, and conservatively assumed that late-type (blue) galaxies do not align, which is consistent with currently available constraints \cite{Mandelbaum11,Heymans13}.
We have shown that the constraints on $A_2$ for future surveys are comparable to, though slightly worse than, current CMB constraints.  Specifically, 
we find $\Delta A_2 = 150 (\tilde b_{\rm NG}^I)^{-1}$ for LSST, while current constraints on $A_2$ from Planck require $-80 < A_2 <110$ at the $68\%$ confidence level.  Note that this constraint was derived for a fiducial $A_2=0$, which neglects any terms scaling as $(A_2)^2$.  To achieve these constraints, imaging surveys will need to access scales below $l<10$, otherwise the constraint worsens by a factor of 1.6.  However, a three-dimensional reconstruction of the auto and cross-spectra of positions and shapes could potentially yield significantly tighter constraints on both $A_2$ and $A_0$, as the alignment process is intrinsically three-dimensional, in contrast to lensing.  Just as importantly, the constraints on $A_2$ from galaxy shapes are to be considered complementary to those from the CMB, as they come from different physical scales ($k\gtrsim 1\,h\,{\rm Mpc}^{-1}$ vs $k\lesssim 0.1 h\,{\rm Mpc}^{-1}$).   

One key caveat to these forecasts is the unknown value of the parameter $b^I_{\rm NG}$, which quantifies the response of galaxy shapes to an anisotropic initial power spectrum.  In this work, we have assumed that $b_{\rm NG}^I$ is of the same order as the tidal alignment strength $b_1^I$ which has been measured observationally. Ideally, $b_{\rm NG}^I$ should be estimated from hydrodynamic simulations of galaxy formation, by running a simulation with an anisotropic initial power spectrum and measuring the response of halos and galaxies to this preferred direction (also as a function of luminosity or mass).  For the purpose of this work, we chose to simply phrase our constraints in terms of the combination $\tilde{b}_{\rm NG}^I A_2$, where $\tilde{b}_{\rm NG}^I \sim 1$ if our assumption is valid.

\acknowledgments
We are thankful to Michael Strauss, Rachel Mandelbaum and Simon White for useful comments and discussions. CD was supported by NASA through Hubble Fellowship grant HST-HF2-51340.001 awarded by the Space Telescope Science Institute, which is operated by the Association of Universities for Research in Astronomy, Inc., for NASA, under contract NAS 5-26555. This work has made use of the Python package mpmath \citep{mpmath}.

\appendix

\section{Derivation of \refeq{3pt}} 
\label{app:calc}

The left-hand side of \refeq{3pt} is given in terms of the
potential three-point function by
\ba
\left\< \d(\vx) \d(\vy) K_{ij}(\vy) \right\> = 
\int \frac{d^3 k}{(2\pi)^3} e^{i\vk\cdot\vr} \M(k) 
\int& \frac{d^3 k_1}{(2\pi)^3} \int \frac{d^3 k_2}{(2\pi)^3} 
\left[\frac{k_{1i} k_{1j}}{k_1^2} - \frac13 \d_{ij}\right]
\vs
& \times
\M(k_1) \M(k_2) \< \phi_{\vk}\phi_{\vk_1}\phi_{\vk_2}\>\,,
\nonumber
\ea
where $\vr = \vx-\vy$, and $\M$ is defined in \refeq{Mdef}.
Using the definition of the bispectrum,
\be
\< \phi_{\vk}\phi_{\vk_1}\phi_{\vk_2}\>
= (2\pi)^3 \d_D(\vk+\vk_1+\vk_2) B_\phi(\vk,\vk_1,\vk_2),
\ee
we have
\ba
\left\< \d(\vx) \d(\vy) K_{ij}(\vy) \right\> =\:& 
\int \frac{d^3 k}{(2\pi)^3} e^{i\vk\cdot\vr} \M(k) 
\int \frac{d^3 k_1}{(2\pi)^3} 
\left[\frac{k_{1i} k_{1j}}{k_1^2} - \frac13 \d_{ij}\right]
\vs
& \times \M(k_1) \M(|\vk+\vk_1|) B_\phi(k,k_1,|\vk+\vk_1|) \, . 
\label{eq:diffterms}
\ea
We now expand the integrand in powers of $q = k/k_1$.  
Further, we define $\mu = \hat\vk\cdot\hat\vk_1$.  
As shown in App.~C of \cite{PBSpaper},
but generalized to the anisotropic case \refeq{BsqI}, the bispectrum is 
given by
\ba
B_\phi(k,k_1,|\vk+\vk_1|) = \sum_\ell A_\ell \P_\ell(\mu) P_\phi(k) P_\phi(k_1) 
\left[ 2 + (2q\mu + [1-2\mu^2]q^2) n_\phi \right] + \O(q^3), \nonumber
\ea
where $n_\phi = n_s-4$ and we have assumed a pure power-law $P_\phi(k)$ 
for simplicity.  Further,
\ba
\M(|\vk + \vk_1|) =\:& \M(k_1)\bigg[ 1 + q\mu \: n_\M(k_1) + \O(q^2) \bigg] ;
\quad 
n_\M(k_1) \equiv \frac{\partial\ln \M(k_1)}{\partial\ln k_1} \,.
\ea
Inserting these expressions into \refeq{diffterms}, we obtain
\ba
&\left\< \d(\vx) \d(\vy) K_{ij}(\vy) \right\> = 
\int \frac{d^3 k}{(2\pi)^3} e^{i\vk\cdot\vr} \M(k) P_\phi(k) \int \frac{k_1^2 dk_1 }{(2\pi)^2} \M^2(k_1) P_\phi(k_1)\vs
&\quad\times  \int_{-1}^1 d\mu \int_0^{2\pi} \frac{d\phi}{2\pi}
\left[\frac{k_{1i} k_{1j}}{k_1^2} - \frac13 \d_{ij}\right] 
\sum_\ell A_\ell \P_\ell(\mu) \left [2 + 2q\mu \left\{n_\phi + n_\M(k_1)\right\} + \O(q^2)\right]\,.
\nonumber
\ea
In the following we will neglect any $\O(q^2)$ corrections; in terms of
observed correlations, these would lead to a small scale-independent correction
which is degenerate with the Gaussian part.    
We now choose $\vk$ to lie along the $z$ axis.  We can then write
\be
\frac{k_{1i} k_{1j}}{k_1^2} = 
\left(\begin{array}{ccc}
\cos^2 \phi (1-\mu^2) & \cos\phi\sin\phi (1-\mu^2) & \cos\phi\: \mu \sqrt{1-\mu^2} \\
\cos\phi\sin\phi (1-\mu^2) & \sin^2 \phi (1-\mu^2) & \sin\phi\: \mu \sqrt{1-\mu^2} \\
\cos\phi\: \mu \sqrt{1-\mu^2} & \sin\phi\: \mu \sqrt{1-\mu^2} & \mu^2
\end{array}\right)\,.
\ee
Performing the $\phi$ integral only leaves the diagonal terms (as expected
by symmetry):
\be
\int_0^{2\pi} \frac{d\phi}{2\pi}
\frac{k_{1i} k_{1j}}{k_1^2} =\frac12 \left(\begin{array}{ccc}
 1-\mu^2 & 0 & 0\\
0 & 1-\mu^2 & 0 \\
0 & 0 & 2\mu^2
\end{array}\right)\,.
\ee
In order to obtain an expression for general directions of $\vk$, we notice
that by symmetry, the final expression can only involve the tensors 
$\hat k^i \hat k^j$ and $\d^{ij}$.  Straightforward algebra then yields
\ba
\int_0^{2\pi} \frac{d\phi}{2\pi}\left[ \frac{k_{1i} k_{1j}}{k_1^2} - \frac13 \d_{ij} \right] = \P_2(\mu) \left[ \frac{k_i k_j}{k^2} - \frac13 \d_{ij}\right]\,.
\ea
We thus have
\ba
\left\< \d(\vx) \d(\vy) K_{ij}(\vy) \right\> =\:&
\int \frac{d^3 k}{(2\pi)^3} e^{i\vk\cdot\vr} \M(k) P_\phi(k) 
\left[ \frac{k_i k_j}{k^2} - \frac13 \d_{ij}\right]
\int \frac{k_1^2 dk_1 }{(2\pi)^2} \M^2(k_1) P_\phi(k_1)\vs
&\times  \int_{-1}^1 d\mu \: \P_2(\mu) 
\sum_\ell A_\ell \P_\ell(\mu) \left [2 + 2q\mu\left\{n_\phi+n_\M(k_1)\right\} + \O(q^2)\right] \vs
=\:& \frac25 A_2
\int \frac{d^3 k}{(2\pi)^3} e^{i\vk\cdot\vr} \M(k) P_\phi(k) 
\left[ \frac{k_i k_j}{k^2} - \frac13 \d_{ij}\right]
\int \frac{k_1^2 dk_1 }{2\pi^2} \M^2(k_1) P_\phi(k_1) \,.
\nonumber
\ea
In the last line, we have used the orthonormality of the Legendre polynomials.  
Note that since $L$ is even, the $q \mu$ term does not contribute and 
corrections are order $q^2$.  The $k_1$ integral now simply gives the
(formally divergent) variance of the density field $\<\d^2\>$.  
By defining
\be
\xi_{\phi \d}(r) = \int \frac{d^3 k}{(2\pi)^3} e^{i\vk\cdot\vr} \M(k) P_\phi(k) \,
\ee
we then obtain
\be
\left\< \d(\vx) \d(\vy) K_{ij}(\vy) \right\> = \frac25 A_2 \D_{ij} \xi_{\phi\d}(r) \<\d^2\>\,.
\label{eq:3ptApp}
\ee
This is \refeq{3pt}.  It is easy to see that the corresponding result
for the quadratic tidal term simply becomes
\be
\left\< \d(\vx) \left[K_{ik} K^k_{\  j} - \frac13 \d_{ij} (K_{lm})^2 \right](\vy) \right\>
= \frac2{15} A_2 \D_{ij} \xi_{\phi\d}(r) \<\d^2\>\,.
\ee
This is because the analog of \refeq{diffterms} for this term differs by factors constructed out of $\vk_{1i}\,(\vk+\vk_1)_j/(k_1|\vk+\vk_1|)$, which at leading order are simply $k_{1i} k_{1j}/k_1^2$.

Also, let us show that anisotropic non-Gaussianity does not lead to a
scale-dependent bias in the \emph{number density} of tracers.  
The tracer density
locally depends on the variance of the density field smoothed on some scale, as well as the square of the tidal tensor $K_{ij}^2$ [\refeq{tijdef}].  
Let us calculate these local quantities given the power spectrum \refeq{Pkloc}:
\ba
\< \d^2 \>_\vx =\:& \int \frac{d^3 \vk_S}{(2\pi)^3} 
\left[1 + f(\hat{\vk}_L\cdot\hat{\vk}_S) \phi(\vk_L) e^{i\vk_L\vx}\right] P(k_S) \vs
\< (K_{ij})^2 \>_\vx =\:& \int \frac{d^3 \vk_S}{(2\pi)^3} 
\left(\hat k_S^i \hat k_S^j - \frac13 \d^{ij} \right)^2
\left[1 + f(\hat{\vk}_L\cdot\hat{\vk}_S) \phi(\vk_L) e^{i\vk_L\vx}\right] P(k_S) \,.
\label{eq:Dphisq}
\ea
Aligning $\vk_L$ with the $z$ axis, we see that both terms involve the integral
\be
\int_{-1}^1 d\mu\: f(\mu) = 2 A_0\,,
\ee
i.e. these quantities pick out the \emph{monopole} of $f$.  Thus, non-Gaussianity of the quadrupolar type does not affect the local averages \refeq{Dphisq} at leading order in the non-Gaussianity.


\section{Renormalized shape biases}
\label{app:newapp}

In this section, we provide a rigorous derivation of the reasoning
outlined in \refsec{IA_renorm}.  For this purpose, we introduce a fictitious
coarse-graining scale $R_L$ which we assume to be much smaller than the scale on which we measure correlations.  
We denote coarse-grained fields by a subscript $L$.  Since we will be dealing exclusively with galaxy shapes and not with number counts, we will drop the superscript $I$ for the remainder of the appendix.

Let us consider the case of Gaussian initial conditions first.  
Since $R_L \gg R_*$, the shape of galaxies at position $\vx$ becomes a local function of the coarse-grained density $\d_L$ and tidal field $K^{ij}_L$.  
In addition, the shapes are in general a functional of the small-scale fluctuations within a region of size $\sim R_*$ around $\vx$.  Hence,
\be
g_{ij}(\vx) = F_{ij}^L(\d_L(\vx) ; K_L^{kl}(\vx); \d_s)\,,
\label{eq:Fij}
\ee
where $\d_s = \d - \d_L$ are the small-scale fluctuations.  The superscript $L$ denotes the fact that $F_{ij}^L$ depends on $R_L$.  
We can now perform a formal Taylor expansion of $F_{ij}^L$ in $\d_L$ and $K_L^{ij}$, taking into account that $F_{ij}^L$ has the same symmetries 
as $g_{ij}$.  The first few terms of the expansion are 
\ba
g_{ij}(\vx) =\:& c_1 K_L^{ij}(\vx)
+ \frac12 c_2 \d_L(\vx) K_L^{ij}(\vx) 
+ \frac12 c_t \left[ K_L^{ik} K_L^{kj} - \frac13 \d^{ij} (K_L^{lm})^2\right]
+ \O(\d_L^3,\, \nabla^2 \d_L)\,.
\label{eq:gammaint}
\ea
The coefficients $c_i$ depend on $R_L$ and on the small-scale modes $\d_s$.  In the Gaussian case, the small-scale modes are uncorrelated with the long-wavelength perturbations, so that they merely lead to stochastic terms which do not correlate over large scales.  Then, the $c_i$ can be regarded as effective constants.  We have not written the most general leading order term here, which would be $c_{ijkl} K_L^{kl}$, and instead reduced $c_{ijkl}$ to a scalar $c_1$.  The reason is that in order to construct anything more general we would need some preferred direction which is absent in 3D space.\footnote{In reality, the line of sight $\vnhat$ provides such a preferred direction.  Thus, in principle a linear term like $\nhat^i \nhat^j \nhat_k \nhat_l t_L^{kl}$ could be present due to redshift space effects.  However such a term will not be relevant for photometric galaxy catalogues.}

The expectation value of $g_{ij}$ has to vanish by symmetry.  Note that this is achieved at second order without having to add any counter term in \refeq{gammaint}.  Now consider the transformation 
\be
K_L^{ij}(\vx,\tau) \to K_L^{ij}(\vx,\tau) + D(\tau) \beta^{ij};\quad
\d_L(\vx,\tau) \to \d_L(\vx,\tau)\,,
\label{eq:betatrans}
\ee
where $D(\tau)$ is the matter growth factor, $\beta^{ij}$ is constant and $\beta_i^{\  i}=0$.   This corresponds to adding a uniform tidal field to the Newtonian
potential,
\be
\Phi_N(\vx) \to \Phi_N(\vx) + \frac34 \Omega_{m0} H_0^2 (1+z) D(z) \beta_{ij} x^i x^j\,.
\ee
Under this transformation, the expectation value of  $g$ changes to
\be
\< g_{ij}(\vx, \tau) \>_\beta = c_1(\tau) \beta_{ij} + \O(\d_L^3,\, \nabla^2 \d_L)\,.
\ee
In the following, we will assume we are evaluating the correlations at some
fixed time $t$, and drop this argument for clarity.  
We now introduce the renormalized bias parameters through
\be
b_1 \equiv \frac{\partial \< g_{ij} \>_\beta}{\partial \beta_{ij}}\Big|_{\beta=0} = c_1 + \O([\d_L,\, K_L^{ij} ]^2)\,.
\label{eq:b1def}
\ee
Note that the definition of $b_1$ is explicitly independent of $R_L$.  The $R_L$-dependent quadratic corrections in \refeq{b1def} come from cubic
order terms in \refeq{gammaint}, hence we have not written them here.  
Similarly, the renormalized quadratic tidal bias is defined via
\be
b_t \equiv \frac{\partial^2 \< g_{ij} \>_\beta}{\partial (\beta_{ij})^2}\Big|_{\beta=0} = c_t + \O([\d_L,\, K_L^{ij} ]^2)\,,
\ee
where now the quadratic corrections come from fourth-order terms in \refeq{gammaint}.

In order to derive the renormalized bias corresponding to $c_2$, we generalize
\refeq{betatrans} to
\be
K_L^{ij}(\vx,\tau) \to K_L^{ij}(\vx,\tau) + D(\tau)\, \beta^{ij};\quad
\d_L(\vx,\tau) \to \d_L(\vx,\tau) + D(\tau)\,c\,,
\label{eq:betatrans2}
\ee
where the transformation of $\d_L$ is the same as the one used to define renormalized local bias parameters in \cite{PBSpaper}.  Specifically, it
corresponds to adding a uniform matter density component $c \rhob$.  We then define
\be
b_2 \equiv \frac{\partial^2 \< g_{ij} \>_{\beta, c}}{\partial \beta_{ij}\partial c}\Big|_{\beta=0, c=0} = c_2 + \O([\d_L,\, K_L^{ij} ]^2)\,.
\label{eq:b2def}
\ee
Again, $b_2$ is independent of $R_L$ by construction.

We now turn to correlations of $g_{ij}$ on large scales.  For
simplicity, let us consider the cross-correlation with matter here, i.e. $\< \d(\vx) g_{ij}(\vy)\>$.  After renormalizing
the $c_n$ into $b_n$ we obtain
\ba
\< \d(\vx) g_{ij}(\vy)\> =\:& b_1 \< \d(\vx) K^{ij}_L(\vy) \>
+ \frac12 b_2 \left\< \d(\vx) \d_L(\vy) K^{ij}_L(\vy) \right\> \vs
&  + \frac12 b_t \left\< \d(\vx) \left[ K_L^{ik} K_L^{kj} - \frac13 \d^{ij} (K_L^{lm})^2\right](\vy) \right\> 
+ \cdots\,,
\label{eq:xidg1A}
\ea
where the ellipsis stands for higher order no-zero-lag correlators, i.e. correlators that do not contain any zero-lag terms such as $\<\d_L^2\>$,
$\<K_{ij} K^{ij} \>$ when expanded into connected correlators.  If $|\vx-\vy| \gg R_L$ (and the correlation function does not have sharp
features), then 
\be
\< \d(\vx) K^{ij}_L(\vy) \> = \< \d(\vx) K^{ij}(\vy) \>
= \D_{ij} \xi(|\vx-\vy|)
\ee
where $\xi(r)$ is the matter correlation function.  This renormalized bias
quantifies the linear response of galaxy shapes to a long-wavelength tidal field as described above.  For Gaussian initial conditions, the three-point
function in \refeq{xidg1A} is only due to non-linear evolution and relevant on smaller scales.

\subsection{Non-Gaussian initial conditions}
\label{sec:NGIA}

We now turn to the case of non-Gaussian initial conditions, quantified by \refeq{BsqI}.  Note that the three-point term in \refeq{xidg1A} 
is sensitive to the bispectrum in the squeezed limit if $|\vx-\vy|\gg R_L$, the coarse-graining scale adopted.  As shown in \refapp{calc}, 
for a squeezed-limit bispectrum of the form \refeq{BsqI} we have 
\ba
\left\< \d(\vx) \d_L(\vy) K_L^{ij}(\vy) \right\>
= \frac25 A_2 \D_{ij} \xi_{\d\phi}(|\vx-\vy|) \s_L^2\,,
\label{eq:3ptA}
\ea
where $\xi_{\d\phi}(r)$ is the matter-potential cross-correlation and $\s_L^2$ is the variance of the coarse-grained density field, $\s_L^2 \equiv \<\d_L^2\>$.  
A similar result (with prefactor $2/15$ instead of $2/5$) holds for the term $\propto b_t$.  
The fact that this term scales as $\s_L^2$, i.e. that
it strongly depends on the arbitrary coarse-graining scale indicates that the bias expansion \refeq{gammaint} is not sufficient in the presence of
primordial non-Gaussianity.  This exactly parallels the case of galaxy density correlations, where the corresponding three-point term also scales
as $\s_L^2$ \cite{PBSpaper}.  

In order to remedy this unsatisfying situation, we explicitly include the dependence of $g_{ij}$ on the small scale fluctuations indicated
in \refeq{Fij}.  Of course, this dependence has to be consistent with the symmetries of $g_{ij}$.  Physically, we expect that the lowest order
dependence on $\d_s$ in \refeq{Fij} will be through a quadrupolar asymmetry in the statistics of the local small-scale density field; an isotropic 
change cannot produce an effect at linear order because we are expanding a spin-2 function.  Note that it is sufficient to phrase the dependence on
the small-scale modes in terms of $\d_s$, since the small-scale tidal field
is related to $\d_s$ through the Poisson equation.  Let us define the parameter $y_s^{ij}$ as the
anisotropy of the local small-scale correlation function within a region of size $R_L$:
\ba
y_s^{kl}(\vx) =\:& \frac1{\s_y^2} \int d^3\v{r}\: W_L(|\v{r}|) \d_s(\vx)  \D_r^{kl} \d_s(\vx+\v{r})
\vs
\s_y^2 =\:& \int \frac{d^3 k_1}{(2\pi)^3} \tilde W_L(k_1) \tilde W_s^2(k_1) P_m(k_1)\,,
\label{eq:ysij}
\ea
where $W_L$ is an arbitrary isotropic filter function on the scale $R_L$
(e.g., Gaussian or tophat), and $\tilde W_L$ is its Fourier transform.  
$\D^{ij}_r$ denotes the derivative operator \refeq{Dijdef} acting with respect to $\v{r}$.  While this definition is convenient, other definitions 
are certainly possible and do not change the results.  Note that since by definition $\d_s = \d-\d_L$, the statistics of $\d_s$ also depend on $R_L$.   
We now let $g_{ij}$ depend \emph{locally} on $y_s^{kl}$:
\ba
g_{ij}(\vx) =\:& F_{ij}^L(\d_L(\vx) ; K^{kl}_L(\vx); y_s^{kl}(\vx))\,.
\ea
We expand $F_{ij}$ to linear order in $y_s^{kl}$, which adds a term
\be
c_{\rm NG} y_s^{ij}(\vx)
\ee
to the expansion on the r.h.s. of \refeq{gammaint}.  At linear order in the non-Gaussianity parameters $A_\ell$, we do not need to consider higher order
terms involving $y_s^{ij}$.  The same symmetry argument as for $K^{ij}_L$ dictates a simple scalar coefficient $c_{\rm NG}$.  
Correspondingly, we have another contribution to the matter-shape
correlation:
\ba
\< \d(\vx) g_{ij}(\vy)\> =\:& b_1 \< \d(\vx) K^{ij}_L(\vy) \>
+ \frac12 b_2 \left\< \d(\vx) \d_L(\vy) K^{ij}_L(\vy) \right\> \vs
&  + \frac12 b_t \left\< \d(\vx) \left[ K_L^{ik} K_L^{kj} - \frac13 \d^{ij} (K_L^{lm})^2\right](\vy) \right\> 
+ c_{\rm NG} \< \d(\vx) y_s^{ij}(\vy) \>
+ \cdots \,.
\label{eq:xidg2}
\ea
The last term can be calculated along the same lines as the calculation
in \refapp{calc}.  The Fourier transform of \refeq{ysij} 
is given by
\be
y_s^{ij}(\vk) = \frac1{\s_y^2} \int \frac{d^3 \vk_1}{(2\pi)^3} 
\left[\frac{k_1^i k_1^j}{k_1^2} - \frac13 \d^{ij} \right] \tilde W_L(k_1)  \d_s(\vk_1) \d_s(\vk-\vk_1)
\label{eq:ysijF}
\ee
We then obtain
\ba
 \< \d(\vx) y_s^{ij}(\vy) \> =\:&
\frac1{\s_y^2} \int \frac{d^3 k}{(2\pi)^3} e^{i\vk\cdot\vr} \M(k) 
\int \frac{d^3 k_1}{(2\pi)^3} \tilde W_L(k_1) \left[k_1^i k_1^j - \frac13 \d^{ij} k_1^2 \right] 
\vs
& \times \M_s(k_1) \M_s(|\vk+\vk_1|) B_\phi(k,k_1,|\vk+\vk_1|) 
\ea
where $\M_s(k) = \M(k) \tilde W_s(k)$.  
The derivation now exactly parallels that of \refeq{3ptApp}, where the $k_1$
integral now yields 
\be
\int \frac{k_1^2 dk_1}{2\pi^2} \tilde W_L(k_1) \M_s^2(k_1) P_\phi(k_1) = \s_y^2\,,
\ee
cancelling the prefactor.  We obtain
\ba
\left\< \d(\vy) y_s^{ij}(\vx) \right\> =\:& \frac25 A_2 \D_{ij} \xi_{\phi\d}(r) \,.
\label{eq:3ptys}
\ea
There are thus three non-Gaussian terms that are proportional to $A_2 \D_{ij} \xi_{\d\phi}$, two of which are strongly dependent on $R_L$.  
Intuition tells us that these contributions should combine to an $R_L$-independent term involving a renormalized bias parameter $b_{\rm NG}$.  
As before, $b_{\rm NG}$ should be given as the response of the mean shape of galaxies to a specific ($R_L$-independent) transformation of the density field.   
Consider the transformation
\be
\d(\vx) \to \d_\alpha(\vx) = \left[1 + \alpha_{lm} \D^{lm} \right]\d(\vx)
\label{eq:alphatransA}
\ee
or equivalently in Fourier space
\be
\d_\alpha(\vk) = \left[1 + \alpha_{lm} \left(\frac{k^l k^m}{k^2}-\frac13 \d^{lm}\right)\right] \d(\vk)\,.
\ee
This corresponds to an anisotropic, scale-independent rescaling of the density field.  In the following we will assume without loss of generality that $\alpha_{lm}$ is trace-free.  The expectation value of the shape after this transformation is
\be
\< g_{ij} \>_\alpha = \frac12 b_2 \< \d_L K_L^{ij} \>_\alpha 
 + \frac12 b_t \left\<  K_L^{ik} K_L^{kj} - \frac13 \d^{ij} (K_L^{lm})^2 \right\>_\alpha 
+ c_{\rm NG}
\< y_s^{ij}\>_\alpha\,.
\ee
where
\ba
\< \d_L K_L^{ij}\>_\alpha =\:&  3 \left\<  K_L^{ik} K_L^{kj} - \frac13 \d^{ij} (K_L^{lm})^2 \right\>_\alpha 
\vs
=\:& \int \frac{d^3k}{(2\pi)^3}
\left(\frac{k^i k^j}{k^2} -\frac13 \d^{ij}\right) \left(1 + \alpha_{lm} \frac{k^l k^m}{k^2}\right) \tilde W_L^2(k) P(k)  \vs
\< y_s^{ij}\>_\alpha =\:& \frac1{\s_y^2} \int \frac{d^3k}{(2\pi)^3}
\left(\frac{k^i k^j}{k^2} -\frac13 \d^{ij}\right) \left(1 + \alpha_{lm} \frac{k^l k^m}{k^2}\right) \tilde W_L(k) \tilde W_s^2(k) P(k)\,.
\ea
After some algebra, this yields
\be
\< g_{ij} \>_\alpha = \left(\frac1{15} b_2 \s_L^2 
+ \frac1{45} b_t \s_L^2
+ \frac2{15} c_{\rm NG}\right)\alpha_{ij}\,.
\ee
The linear response of the mean 
shape of galaxies, which shall define our bias $b_{\rm NG}$, is thus given by
\be
b_{\rm NG} \equiv \frac{\partial \< g_{ij} \>_\alpha}{\partial\alpha_{ij}}\Big|_{\alpha=0} = \frac1{15} b_2 \s_L^2 
+ \frac1{45} b_t \s_L^2
+ \frac2{15} c_{\rm NG}\,.
\label{eq:b01}
\ee 
Solving this for $c_{\rm NG}$ yields 
\be
c_{\rm NG} = \frac{15}2 b_{\rm NG} - \frac12 b_2 \s_L^2 -\frac16 b_t \s_L^2\,,  
\ee
which we insert into \refeq{xidg2} to obtain
\ba
\< \d(\vx) g_{ij}(\vy)\> 
=\:& b_1 \D_{ij} \xi_{\d\d}(|\vx-\vy|) + 3 b_{\rm NG} A_2 \D_{ij} \xi_{\d\phi}(|\vx-\vy|)\,.
\label{eq:xidgrenormA}
\ea  
Our intuition is thus confirmed: by adding the bias $b_{\rm NG}$, which
quantifies the linear response of galaxy shapes to an anisotropic power spectrum of initial fluctuations, we have absorbed the unphysical dependence on
$\sigma_L^2$ present in the local quadratic bias expression.

\section{Non-Gaussian contributions to weak lensing shear}
\label{app:shearNG}

In \refsec{Cl} and following, we have neglected any modification of the
lensing contribution to galaxy shapes due to primordial non-Gaussianity.  
We now show that there is a modification, but it is negligible compared to
the scale-dependent signature in the intrinsic alignments.  The
observed shear is not exactly given by its linear expression $\gamma$ 
introduced in \refsec{Cl}, but rather there are nonlinear corrections,
\ba
\gobs(\vth) =\:& \g(\vth) + c_\g \k(\vth) \g(\vth) + \cdots \label{eq:gobs}\,,
\ea
where $c_\g$ is a constant of order unity.  The nonlinear corrections are
due to reduced shear \cite{Bartelmann01} as well as lensing bias corrections
induced by source density weighting \cite{lensingbias}.  
This leads to a three-point contribution to shear two-point functions
of the form
\ba
\<\gobs(\vth)\gobs(\vth')\> =\:& \<\g(\vth)\g(\vth')\> 
 + 2 c_\g \<\k(\vth)\g(\vth)\g(\vth')\> + \cdots\,.
\label{eq:shearcorr}
\ea
Primordial non-Gaussianity in general contributes to this three-point
function.  This was studied in detail by Ref.~\cite{shearfNL}, who showed that  
\ba
C_{\gobs}(l) =\:& C_\g(l) 
+ 2 c_\g \!\int\!\frac{d^2l_1}{(2\pi)^2}
\cos 2\phi_{l_1} B_\k(l_1,|\v{l}-\v{l}_1|,l)\label{eq:dCg}\,.
\ea
Here, $B_\k$ is the convergence bispectrum, and we have aligned $\v{l}_1$ so that $\phi_l=0$.  Further, we have used that in Fourier space $\g(\v{l}) = \g_1(\v{l}) + i \g_2(\v{l}) = \exp(2i\phi_l) \k(\v{l})$.  Assuming the Limber approximation, the convergence bispectrum is related to the matter bispectrum by
\bea
B_\k(l_1,l_2,l_3) &=& \left(\frac32 \Om H_0^2\right)^3 \int_0^{\chi_s}
\frac{d\chi}{\chi} \frac{\mathcal{W}_{\rm L}^3(\chi_s,\chi)}{\chi^3\:a^3(\chi)}
B_m \!\left (\frac{l_1}{\chi},\frac{l_2}{\chi},\frac{l_3}{\chi}; \chi \right ),
\eea
where $\mathcal{W}_{\rm L}(\chi_s,\chi)\equiv \chi/\chi_s (\chi_s-\chi)$ denotes the lensing kernel.  
Ref.~\cite{shearfNL} only considered isotropic local non-Gaussianity, however their results are easily generalized to the anisotropic case by making use of the squeezed limit.  While in the isotropic case, the low-$l$ contribution to
\refeq{dCg} is canceled by the $\cos2\phi_{l_1}$ modulation, this factor will lead to a non-zero contribution in the anisotropic case, which will be of the same order as the isotropic contribution to $C_{\kobs}(l)$ discussed in \cite{shearfNL}.  

For our purposes, an order of magnitude estimate suffices.  By using the squeezed limit for the bispectrum and approximating the projection kernels, we can derive in analogy with Sec.~II in \cite{shearfNL}
\be
\frac{C_{\gobs}(l)}{C_\g(l)} - 1\sim 10^{-5} \left(\frac{A_2}{100}\right) 
\left(\frac{l}{100}\right)^{-2}\,.
\label{eq:oom}
\ee
Thus, at the values for $A_2$ relevant for detectable alignment signatures,
the fractional modification to the shear is much less than one even at the lowest multipoles.

\bibliographystyle{unsrt}
\bibliography{PBSng}

\end{document}